\documentclass[10pt,journal,twocolumn]{IEEEtran}
\usepackage{cite}
\usepackage{multirow} 
\usepackage{algpseudocode}
\usepackage{algorithm}
\usepackage{rotating}
\usepackage[normalem]{ulem}

\usepackage{kantlipsum} 
\usepackage{commath}
\allowdisplaybreaks
\usepackage{mathtools}  
\usepackage{verbatim}
\usepackage{xr-hyper} 
\usepackage{enumitem}

\usepackage{epstopdf}
\usepackage{wrapfig}
\usepackage{latexsym}
\usepackage{amssymb}
\usepackage{amsthm}
\usepackage{amsfonts}
\usepackage{amsmath} 
\usepackage{graphicx}
\usepackage{latexsym}
\usepackage{booktabs}
\usepackage{subcaption} 

\usepackage[T1]{fontenc}

\usepackage{breqn}

\algnewcommand\algorithmicinput{\textbf{INPUT:}}
\algnewcommand\INPUT{\item[\algorithmicinput]}
\algnewcommand\algorithmicoutput{\textbf{OUTPUT:}}
\algnewcommand\OUTPUT{\item[\algorithmicoutput]}
\usepackage[dvipsnames]{xcolor}
\colorlet{BLUE}{blue}
\colorlet{ROYALPURPLE}{RoyalPurple}
\colorlet{RED}{red}

\newif\ifcommentson
\commentsontrue

\newif\ifextended
\newif\ifshortver

\shortvertrue

\newcommand{\extended}[1]{\ifextended \ifshortver \textcolor{red}{#1} \else \textcolor{black}{#1} \fi  \fi}
\newcommand{\shortver}[1]{\ifshortver \ifextended \textcolor{blue}{#1} \else \textcolor{black}{#1} \fi \fi}

\newcommand{\optional}[1]{\ignorespaces}
\begin{document}
\bstctlcite{IEEEexample:BSTcontrol}

\title{SDN Architecture and Southbound APIs for IPv6\\Segment Routing Enabled Wide Area Networks}

\author{Pier Luigi Ventre, Mohammad Mahdi Tajiki, Stefano Salsano, Clarence Filsfils
\IEEEcompsocitemizethanks{\protect
\IEEEcompsocthanksitem P.L. Ventre and S. Salsano are with the Department of Electronic Engineering at the University of Rome ``Tor Vergata'' and the Consorzio Nazionale Interuniversitario per le Telecomunicazioni (CNIT) - Rome, Italy E-mail: \{pier.luigi.ventre, stefano.salsano\}@uniroma2.it
\IEEEcompsocthanksitem M.M. Tajiki is with the Department of Electronic and Computer Engineering at the University of Tarbiat Modares - Tehran, Iran E-mail: mahdi.tajiki@modares.ac.ir
\IEEEcompsocthanksitem C. Filsfils is with Cisco Systems - Paris, France E-mail: cfilsfil@cisco.com
}}

\markboth{\textbf{Submitted to IEEE Transactions on Network and Service Management}}%
{Ventre \MakeLowercase{\textit{et al.}}: SDN Architecture and Southbound APIs for IPv6 Segment Routing Enabled Wide Area Networks}

\IEEEtitleabstractindextext{
\begin{abstract}
The SRv6 architecture (Segment Routing based on IPv6 data plane) is a promising solution to support services like Traffic Engineering, Service Function Chaining and Virtual Private Networks in IPv6 backbones and datacenters. The SRv6 architecture has interesting scalability properties as it reduces the amount of state information that needs to be configured in the nodes to support the network services. In this paper, we describe the advantages of complementing the SRv6 technology with an SDN based approach in backbone networks. We discuss the architecture of a SRv6 enabled network based on Linux nodes. In addition, we present the design and implementation of the Southbound API between the SDN controller and the SRv6 device. We have defined a data-model and four different implementations of the API, respectively based on gRPC, REST, NETCONF and remote Command Line Interface (CLI). Since it is important to support both the development and testing aspects we have realized an Intent based emulation system to build realistic and reproducible experiments. This collection of tools automate most of the configuration aspects relieving the experimenter from a significant effort. Finally, we have realized an evaluation of some performance aspects of our architecture and of the different variants of the Southbound APIs and we have analyzed the effects of the configuration updates in the SRv6 enabled nodes.
\end{abstract}

}
\maketitle
\IEEEdisplaynontitleabstractindextext
\IEEEpeerreviewmaketitle

\vspace{-1em}

\section{Introduction}
\label{sec:intro}

\IEEEPARstart{T}{he} Segment Routing (SR) architecture \cite{idsrarch}\cite{filsfils2015segment} gives the possibility to include a list of instructions (called \textit{segments}) in the packet headers. This \textit{Segment List} influences the forwarding path of the packets and can also provide \textit{instructions} to be performed on a packet in a given node. The Segment Routing architecture can support several use cases of great value for Service Providers, like: Traffic Engineering, Service Function Chaining (SFC), Fast Failover, Operation And Management (OAM), Virtual Private Networks (VPNs). For a more exhaustive list of use cases see~\cite{segment-routing.net}.

There are two variants of the Segment Routing architecture, as it can be implemented using either the MPLS or the IPv6 data plane for packet forwarding. In the former case, the Segment IDs (SID) are expressed using MPLS labels and a \textit{Segment List} is a ``stack'' of labels in the MPLS header. In the latter case, the SIDs are expressed using IPv6 addresses and the \textit{Segment List} is carried in a new type of IPv6 Routing Extension Header called SR Header (SRH)~\cite{idipv6srh}. \extended{Considering the huge IPv6 addressing space, it is easier to encode \textit{instructions} and not only \textit{locations} in an IPv6 SID. In~\cite{srv6netprog}, the IPv6 Segment Routing concept is extended from the simple steering of packets across nodes to a general \textit{network programming} approach.} In this paper we focus on the IPv6 data plane and we will refer to the Segment Routing implemented on the IPv6 data plane as \textit{SRv6}.   

In general, the advantage of Segment Routing is the possibility to add state information in the packet headers, avoiding or minimizing the information to be configured in the internal nodes to realize network services. Adding state information in the packets at the network edge, as opposed to reconfigure internal network nodes, greatly improves the scalability of services based on SR and allows simpler and faster service setup and reconfiguration. In particular with SRv6 we can use a connection-less forwarding technology like IPv6 and obtain the same flexibility and degree of control of a technology like MPLS. For the above considerations, the SRv6 technology is attracting interest from Service Providers and equipment vendors \cite{SRv6blog-noction}. In \cite{segment-routing.net}, the documents under discussion in IETF and the open source initiatives for implementing SRv6 are reported. Noteworthy, Segment Routing support has been included in the recent Linux kernels and in the VPP platform developed by the open source initiative FD.io \cite{fd-io-vpp}.

The Software Defined Networking (SDN) concept \cite{mckeown2009software,kreutz2015software} is now becoming ubiquitously widespread both in data center networks and in large scale wide area networks. In the original SDN concept, the control plane (i.e. the network intelligence) is logically centralized and separated from the data plane. The \textit{SDN controller} is the entity which implements the control plane functionality and takes full control of the forwarding devices running in the data plane. The communication between these two layers is handled via an interface called Southbound API. Currently, the SDN concept is mostly used in a wider meaning, also considering the remote configuration of arbitrary devices that include their own control plane. Therefore SDN becomes almost synonymous of automatic and centralized configuration.

It is natural to consider an SDN based approach to control Segment Routing based services in a Service Provider network. A centralized logic can take decisions concerning the Segment Lists that need to be applied to implement the services, then the \textit{SDN controller} can interact with the edge nodes to enforce the application of such Segment Lists. The possibility for the SDN controllers to interact only with edge nodes to setup and reconfigure complex services is extremely appealing from the point of view of the simplicity and efficiency of the solution. The centralized vision of the \textit{SDN controller} can be used to perform an \textit{optimal} selection of the Segment Lists to be applied to the packets flows. Different optimality criteria can be considered, e.g., share equally the load in the network, reduce the energy consumption, maximize the resilience to faults, in any case these criteria are out of the scope of this work. 

In this paper, we present the design and implementation of a SDN architecture to control SRv6 enabled networks. This work has been performed in the context of the ROSE research project \cite{rose} which aims at creating an open source SRv6 ecosystem. In particular,  we designed and implemented a Linux based SRv6 node made of open source components. The proposed SRv6 node exposes an API towards the SDN controller, which is a \textit{Southbound} API considering it from the perspective of the \textit{SDN controller}. Therefore we will refer to it as the SRv6 Southbound API. We have focused on the definition and implementation of the SRv6 Southbound API considering 4 different variants (gRPC, REST, NETCONF, SSH/CLI). We have released the implementation of the Linux SRv6 based node, the specifications of the SRv6 Southbound API and their implementation as open source libraries (written in python) for the \textit{SDN controller} and as modules to be installed in the Linux based routers \cite{srv6-sdn}. Our contributions also include a set of management tools which offer an Intent based API to the users and allow to realize replicable testbeds on Mininet emulator \cite{mininet}, distributed IaaS infrastructures and physical testbeds \cite{rose}. Using these tools we have realized: i) a performance evaluation and a comparison of the different implementations of the Southbound APIs; ii) an analysis of the effects of the SRv6 configuration changes in the Linux SRv6 nodes, showing that we can achieve hitless reconfiguration of SRv6 policies.

The paper is structured as follows: in Section~\ref{sec:SRv6intro} we provide a short introduction on the SRv6 technology. Section~\ref{sec:SRv6node} presents the architecture of the Linux SRv6 node, followed by the SDN architecture in Section~\ref{sec:SDNarch}. The details of the SRv6 Southbound API are discussed in Section~\ref{sec:southboundAPI}. Section~\ref{sec:emulation} describes the emulation tools that we have released, while the experimental results are illustrated in Section~\ref{sec:experimental}.

\section{IPv6 Segment Routing (SRv6)}
\label{sec:SRv6intro}

The Segment Routing (SR) architecture \cite{idsrarch}\cite{filsfils2015segment} is based on (loose) source routing. Basically, it allows the source of a packet (e.g. a host or a router) to add a list of \textit{Segments} to a packet header. According to~\cite{filsfils2015segment}, \textit{a Segment is an identifier for a topological instruction (steering the packet over a given path) or a service instruction (delivering the packet to a service)}. The SR architecture can run over a MPLS or an IPv6 data plane. \shortver{It supports several use cases of great interest for a Service Provider, like for example Traffic Engineering, Network Resilience and VPNs (see \cite{RFC7855} \cite{RFC8355} \cite{RFC8354} \cite{id-ti-frr-sr}).}\extended{It supports several use cases of great interest for a Service Provider, as an example it can be used to support a Fast Reroute solution called TI-LFA (Topology Independent-Loop Free Alternate)~\cite{id-ti-frr-sr}. Other examples are SFC and VPNs. Segment Routing Header (SRH) is inline with the standardized SFC architecture \cite{RFC7665}. The most important difference with respect to the other implementations is that there is no need to have state in the SF forwarders. Moreover, SRv6 allows to implement SFC in the network without the need of having completely separated \textit{overlay} and \textit{underlay} operations. As regards the VPNs, the ingress edge node encapsulates the customer edge packets in an outer IPv6 header where the destination address is the SRv6-VPN SID provided by the egress node. The underlay between the edge nodes only needs to support plain IPv6 forwarding and there is no need of encapsulation protocols such as L2TP, VXLAN and GRE. In \cite{srforsdwan} it is explained how SR enables underlay Service Level Agreements (SLA) to a VPN with scale and security while ensuring service opacity.} Moreover, it has been designed to be high scalable as explained in \cite{interconnecting} where the scaling capability of Segment Routing has been demonstrated considering an use case of 600,000 nodes and 300 millions of endpoints.

The concepts of SRv6 have been extended in \cite{srv6netprog}: each Segment represents not only a location but also a function to be called at a specific location in the network. A function can represent a simple action like forwarding or a complex behavior defined by the user. Each SRv6 capable node maintains the ``My Local SID Table'' where the association of SIDs with these local functions is defined. In order to signal the availability of a function, a node can advertise it using an IGP routing protocol leveraging the fact the SIDs are represented as regular IPv6 addresses (this is also an advantage with the respect of SR-MPLS which requires extensions to the routing protocols). Combining these ``network instructions'' it is possible to literally program the networks and realize very complex behaviors minimizing the need of state information in the core network nodes. 

In this work, we focus on the IPv6 data plane solution for Segment Routing, in short SRv6. In this solution, the \textit{Segment List} (or \textit{SID List}) is carried in the Segment Routing Header (SRH)~\cite{idipv6srh}. Three basic operations are defined with respect to the SRH: encapsulation, processing, and decapsulation. The SRH added by the source (encap operation) contains a number of ``intermediate'' SIDs, the final destination of the packet, and a pointer called ``Segments Left'' (SL) which points to the ``active'' SID, i.e. the next SID to be processed. The SR information can be pushed into the packets using two different approaches, denoted as \textit{insert} and \textit{encap} modes, respectively. When a node uses the \textit{insert} mode the SRH is pushed as next header in the original IPv6 packet, immediately after the IPv6 header and before the transport header. The original IPv6 header is changed, in particular the \textit{next header} is modified according to the value of SRH, the IPv6 destination address is replaced with the IPv6 address of the first SID in the \textit{Segment List}, while the original IPv6 destination address is carried into the SRH header as the last segment of the list. In the \textit{encap} mode, an outer IPv6 header is pushed, which carries the SRH header with the \textit{Segment List}. While the original IPv6 packet is transported as the inner packet of an IPv6-in-IPv6 encapsulated packet and travels unmodified in the network.

Let us consider a packet sent by a node A1, with a SID list that contains two intermediate SIDs (A2 and A3) and the IPv6 destination address (A4). In this case the SL pointer will start at 2 and the packet will be sent by A1 by setting the IPv6 destination address to A2. When A2 receives the packet, it decrements the SL pointer to 1, so that the next SID becomes A3, and it copies A3 into the IPv6 destination address. Note that the SID list is not modified, only the SL pointer is decremented. When A3 receives the packets it decrements SL to 0 so that it points to A4 (the final destination) and copies A4 into the IPv6 destination address.

\section{Linux SRv6 node architecture}
\label{sec:SRv6node}

We propose the architecture of a Linux based SRv6 node, which foresees the coexistence of a local control logic based on distributed IP routing and of an SDN approach in which the node offers an API towards an \textit{SDN controller}. This is a \textit{Southbound API} as seen from the controller point of view. Similar solutions have been proposed recently and often referred to as \textit{hybrid} IP/SDN (see for example \cite{salsano2016hybrid}). The peculiarity and novelty of our approach is that we consider an SRv6 enabled network, hence our node offers a \textit{SRv6 Southbound API}.

\begin{figure}
    \centering
    \includegraphics[width=0.45\textwidth]{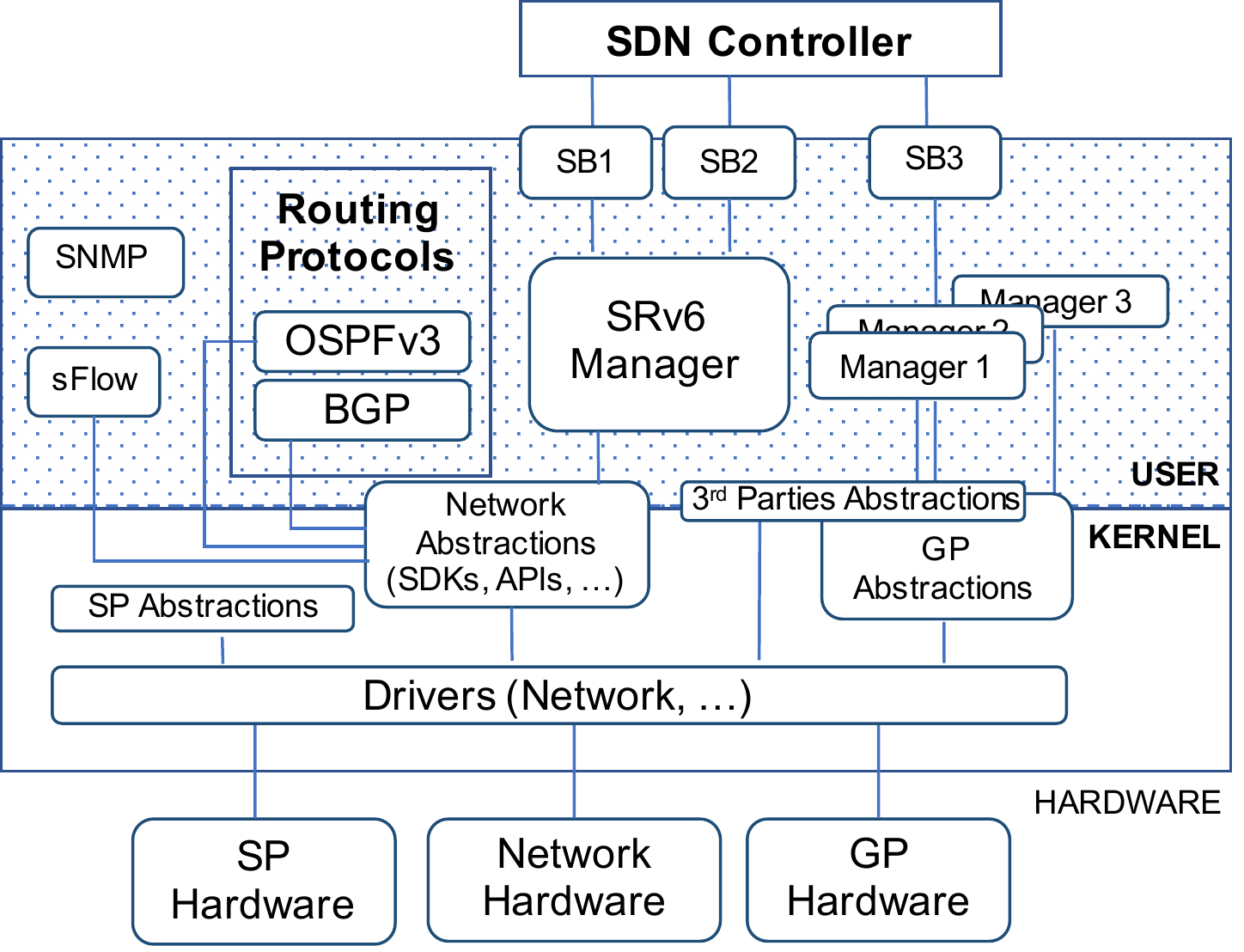}
    \caption{Linux based SRv6 node architecture}
    \label{fig:srv6_node}
    \vspace{-3ex}
\end{figure}

Figure \ref{fig:srv6_node} shows the high level architecture of our Linux based SRv6 node. This logical architecture can be specialized and applied to different physical realizations, ranging from specialized hardware boxes to general purpose servers. We discuss Figure~\ref{fig:srv6_node} proceeding bottom up.

At the lowest level, there are the hardware resources that we have classified in \textit{Special Purpose}, \textit{Network} and \textit{General Purpose} resources. Considering a specialized network node, the hardware resources would be the Switching Silicon (ASIC) and possibly other peripherals. On top of this, we envisage a Linux based Operating System (OS), which could be either a general purpose Linux distribution or a specialized one like Open Network Linux (ONL) \cite{onl}. ONL is a Linux distribution made for open hardware switches, and in general network devices built from commodity hardware.

The operations needed to control the hardware resources (that perform data plane packet processing) are performed in the kernel space. The Linux OS is shipped with the necessary drivers which allows the kernel space to use the hardware resources. On top of this, there are the kernel abstractions and their APIs; they are provided in the form of kernel modules and include also what is necessary to directly program the network resources, to properly use GP hardware and so on. In this level we find the \textit{Network SDKs} which offers proper means to access the resources of the specialized nodes like switches or routers. There is already a number of devices available in the market that follow this design approach. For example a list of hardware switches certified for Open Networking Linux can be found at~\cite{onl}. Cumulus Linux~\cite{cumulus} is another example of a customized version of Linux that is used by Cumulus for their Linux based solutions.

IPC (Inter Process Communications) mechanisms like the netlink protocol \cite{rfc3549} represents the way through which user and kernel spaces communicate. In the user space level, we foresee the coexistence between the control logic based on distributed routing control protocols and the SDN approach: protocols like OSPFv3 \cite{rfc5340} and BGP \cite{rfc2545} program the \textit{Network Abstractions} using their internal logic. The IP routing protocol allows the nodes to exchange the basic reachability information (IPv6 prefixes) about all network entities including the IPv6 addresses that will be used as Segment Identifiers (SIDs). The decisions taken by the routing protocols can be overridden by a \textit{SDN controller} which programs the SRv6 instructions in the nodes leveraging the \textit{Southbound API} exposed by the network itself. The main component of this architecture is the so called \textit{SRv6 Manager} which acts as mediators interacting on the south with the SRv6 abstractions offered by the kernel and on the north with the \textit{SDN controller}. \extended{This approach can be compared with the traditional SDN solutions based for example on the OpenFlow (OF) protocol \cite{mckeown2008openflow}. In these solutions, an OF based switch includes an user space agent translating the OF messages into actions to be submitted to the kernel components (see for example the OF-DPA software architecture for the control of the switches based on the Broadcom chipset~\cite{ofdpa})} In our architecture the \textit{SRv6 Manager} is the user space agent translating the messages received over the \textit{SRv6 Southbound API} into actions to be sent to the kernel components.

\subsection{Implementation of the node architecture}

\begin{figure}
    \centering
    \includegraphics[width=0.40\textwidth]{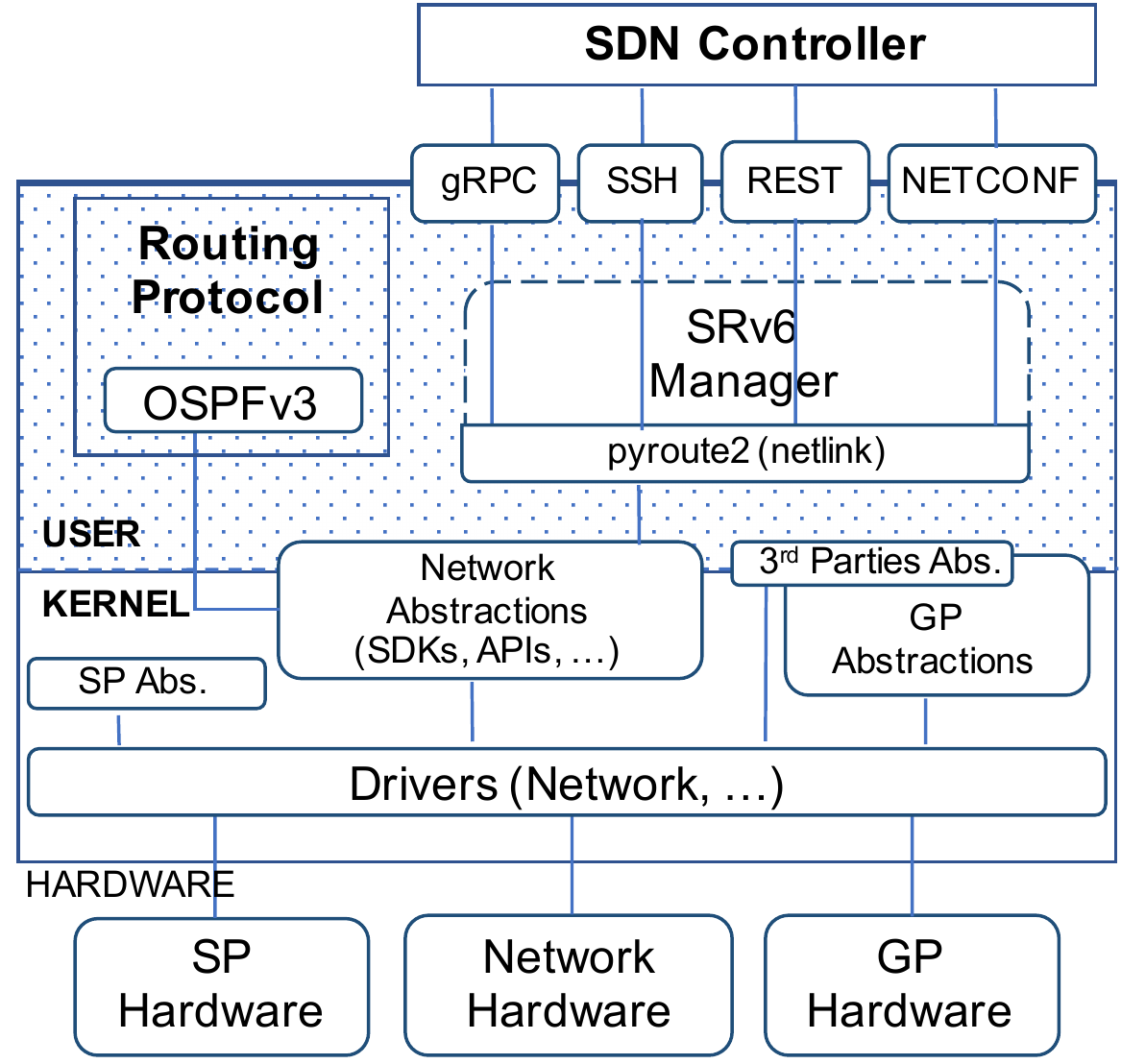}
    \caption{Linux SRv6 node: implemented components}
    \label{fig:srv6_node_impl}
    \vspace{-3ex}
\end{figure}

We have implemented a Linux SRv6 node leveraging commodity hardware. We have used a general purpose distribution of the Linux OS, we only require the kernel to be recent (at least 4.10) in order to have native support for SRv6 operations in the kernel space. For our purposes, we did not need to enhance or modify the existing Linux kernel support of SRv6 \cite{lebrun2017implementing}, we only worked on the the user space components necessary to implement the previously described node architecture. In particular, we focused on the \textit{SRv6 Manager} component and on the implementation of \textit{SRv6 Southbound API}. Figure~\ref{fig:srv6_node_impl} shows the software modules included in our node implementation. 

The \textit{SRv6 Manager} component is developed in python. It allows to translate the instructions carried by the Southbound protocols into SRv6 instructions which are submitted to the Linux kernel. As shown in Figure~\ref{fig:srv6_node_impl}, the \textit{SRv6 Manager} can support four different variants of the SDN \textit{Southbound API} (gRPC, SSH/CLI, REST and NETCONF), as it will be discussed in Section~\ref{sec:propSouthAPI}. The communication among the \textit{SRv6 Manager} and the Linux kernel is based on the open source project pyroute2 \cite{pyroute2}, a pure python \textit{netlink} library. We have added the support for the SRv6 functionality and this contribution has been accepted and merged in the mainstream distribution of pyroute2. In this way, the messages coming from the \textit{SDN controller} can be translated directly in \textit{netlink} messages. In the future, we plan to extend the \textit{SRv6 Manager} taking advantages of the other protocols managed by pyroute2. For example, we could inform the \textit{SDN controller} of the interfaces going up/down, or we can bring administratively up/down a network interface. In other words, the same architecture that we have used to control the SRv6 capabilities can be reused to control all other types of networking capabilities of a Linux node.

In the implemented Linux SRv6 node the distributed routing protocol component is an OSPFv3 daemon. We have leveraged the open source implementation provided by the Quagga Routing suite~\cite{quagga}. The OSPFv3 daemon distributes the IPv6 reachability information setting up the basic connectivity among all network entities, in this way the IP forwarding can be always used as default forwarding. We have also used OSPFv3 to feed the \textit{SDN controller} with topological information about the SRv6 enabled network (see Section \ref{sec:TopoDisc} for further details).

\section{SDN architecture for SRv6 WANs}
\label{sec:SDNarch}

We consider a Service Provider offering network services based on SRv6. The edge and core IPv6 routers of the Service Provider network constitute a SRv6 domain. Customer packets entering in the edge routers are classified and encapsulated into IPv6 packets that can include a Segment Routing Header with the Segments List. While the SRv6 domain is operating with IPv6, there are no limitations on the type of customer traffic that can be supported: IPv4 and IPv6 transport services can be offered, as well as pure layer 2 connectivity services. The IPv6 routing information needed for the forwarding of packets in the SRv6 domain is exchanged among edge and core routers using well established IGP routing protocols.

Differently from a classic SDN approach based on OpenFlow, the SDN controller does not need to interact with all the edge and core nodes to discover the network topology and to setup the packet forwarding rules. In our SDN architecture for SRv6 enabled WANs, as for the topology discovery we assume that the SDN controller can interact with routers and synchronize with the vision of the topology computed by the routing protocol (see Section \ref{sec:TopoDisc} for further details). As regards the setup of forwarding rules, it is possible for the SDN controller to interact with a single node (the ingress edge node) to enforce the application of a Segments List to a given flow (see Section \ref{sec:SetupSRFlow}). Of course, this architecture does not preclude the possibility for the SDN controller to interact with the core nodes for other services and use cases. 

SDN approaches fully based on OpenFlow have been demonstrated to work well in data-center scenarios, where all the switches to be controlled are reachable with sub-millisecond delay. Controlling a WAN in which the network latency between the SDN controller and the forwarding nodes can be in the order of hundreds of milliseconds with the same approach is a much more challenging task. For this reason we advocate that even in a Software Defined WAN scenario it is better to manage the basic connectivity using traditional distributed routing protocols. These solutions can be augmented with Fast Reroute mechanisms~\cite{id-ti-frr-sr}, which in turn could rely on an SDN approach for their configuration.

\subsection{Topology discovery}
\label{sec:TopoDisc}

The SDN controller will be running in a datacenter or in a point of presence, usually co-located with one or more routers belonging to the SRv6 domain. Assuming that a link-state IGP routing protocol is running in the SRv6 domain, we want to discuss how the SDN controller can become aware of the network topology and of its changes. A fundamental characteristic of link state routing protocols is that each router becomes aware of the full network topology in its topological database. In this discussion, we refer to the specific routing protocol (OSPFv3) that we have integrated in our implementation and also on the specific open source implementation for Linux that we have used. Nevertheless, our considerations have a more general architectural validity for other link-state routing protocols. 

We propose two different approaches to let the SDN controller become aware of the topology, that we call respectively \textit{TI-Extraction} (Topology Information Extraction) and \textit{TD-Entity} (Topology Discovery Entity). In the \textit{TI-Extraction} case only the routers take part in the (OSPFv3) routing protocol exchanges, the topological information is extracted from one or more routers and transferred to the SDN controller. From the implementation point of view, the simplest approach is to log in into a router using the CLI (Command Line Interface), dump the topological database and parse it to extract the topology to be sent to the controller. This could be achieved with a periodic polling. A relatively short polling interval (e.g. in the order of few seconds) could be used to promptly update the topology in the SDN controller. Update operations at a time scale of few seconds are acceptable, as we recall that the fast reaction to link/node failures is not meant to be realized through an immediate reaction by the SDN controller. A more efficient approach is to implement a software module integrated in the router that can export the topology. For example considering the Linux platform, the Quagga OSPFv3 daemon could be enhanced to support this feature and interact with a SDN controller. In this case, the SDN controller could be notified of the topology changes by the enhanced routing daemon rather than using a periodic polling.

In the \textit{TD-Entity} case, a Topology Discovery Entity interacts with one or more routers using the (OSPFv3) routing protocol, with the only purpose of building and updating the topological database. The \textit{TD-Entity} can be integrated in the SDN controller or it can be a separate entity that communicate the topology to the SDN controller in some way. For example Quagga offers the so called Quagga Fib Push Interface through which the daemon can notify the learned routes to an external entity. This can be the SRv6 Manager running in the node or directly the SDN controller. In the existing implementations, it is possible find also solutions integrating OSPFv3 speakers in the Southbound of the SDN controllers. This approach has the advantage of not having to modify/enhance a specific implementation of the routing protocol and provides the same advantages in terms of readiness without any polling procedure. The fact that the Topology Discovery Entity can be connected to more than one router is very useful from the resiliency point of view, as the SDN controller can remain updated on the network topology status also in case of failures, as long as there is at least one router which remains active and connected to the SDN controller. 

\begin{figure}
    \centering
    \includegraphics[trim={11cm 8.5cm 11.5cm 1cm},clip,width=0.45\textwidth]{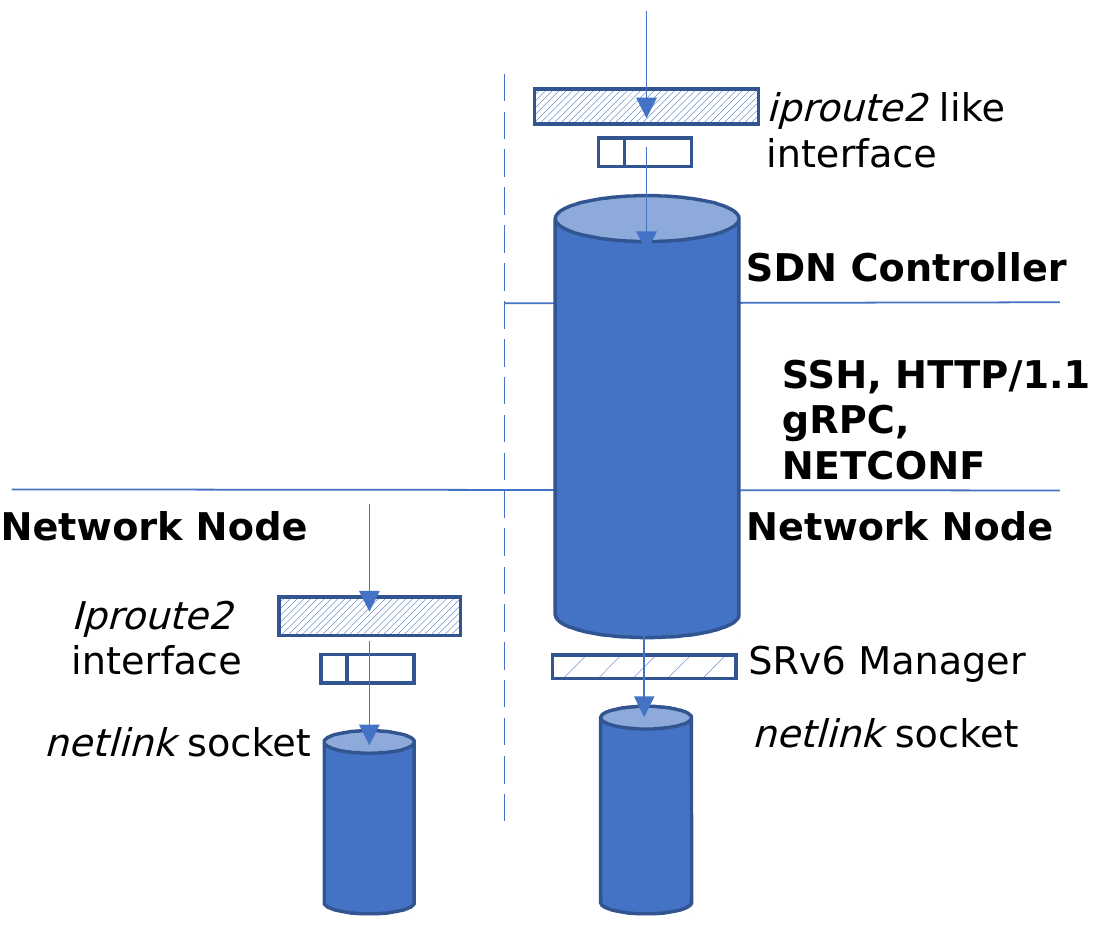}
    \caption{SRv6 Southbound API design}
    \label{fig:srv6_sb}
    \vspace{-3ex}
\end{figure}

For our implementation, we have followed the \textit{TI-Extraction} approach. Our entity is actually a process, running alongside with the SDN controller, which connects to a given router and dumps its topological database and build a network graph using the extracted information.

\subsection{Setup of SR based services}
\label{sec:SetupSRFlow}

The setup of SR based services consists in the configuration of ingress edge routers to classify the incoming packets, associate them with the proper Segments List and SRv6 behavior. Moreover it may be needed to associate further SRv6 behaviors to the Segment Identifiers (SIDs), both in edge and in core routers. 

In the proposed SDN based architecture, the SDN controller is in charge of performing these configuration operations. From the point of view of the SDN controller, the SRv6 Southbound API represents the functionality that can be offered by the SRv6 node and used by the SDN controller to setup the SR based services. We considered the SRv6 functionality offered by a Linux router in the most recent kernel versions and we carried out a thorough analysis in order to identify the requirements from the perspective of the Southbound API. The main functional requirements we have identified are the support of the L3 transit behaviors \cite{srv6netprog}, which include \textit{encap} mode for IPv4 and IPv6 traffic and \textit{insert} mode for IPv6 traffic. The creation of a SR policy inside the node and association to a supported behavior. The removal of a specific SR policy; the update of a SR policy which can affect the Segments List, the behavior and the security mechanisms. Finally, listing of the policies installed in the nodes. In future, we plan to extend this SRv6 Southbound API to support the functionalities described in \cite{srv6netprog}. More in detail, the SRv6 operations that are available in a Linux node can be accessed through \textit{iproute2}, a collection of utilities which allow to control and monitor most of the aspects of the networking in the Linux kernel. \textit{iproute2} is a user space utility that offers a CLI and uses a \textit{netlink} socket to communicate with the kernel. We consider the SRv6 commands and corresponding parameters available in \textit{iproute2} as the reference to define the SRv6 Southbound API. Figure \ref{fig:srv6_sb} provides a visual representation of the design choices behind the implementation of the proposed APIs. With this approach, we expose on the SRv6 Southbound API the same functionality that a local application running inside the SRv6 Linux node would have.

It is worth to notice that is important to define not only the functionality offered by this Southbound API but also the protocol mechanisms at its ground. The protocol mechanism should be lightweight, that is the processing overhead should be reasonably low, since the agent managing the protocol should coexist also with the IP routing protocols. It should be network efficient in terms of traffic exchanged and response time to perform an operation. It is important to take into account also the robustness to network impairments like packet loss and delay. Finally, we do believe that the protocol mechanism should support proper security which means at least authentication of the parties involved in the communication, protection of the privacy and integrity of the exchanged data. Insecure mechanisms have been considered in Section \ref{sec:experimental} only for benchmarking purposes,

As illustrated in Section \ref{sec:southboundAPI}, there is a large number of different technologies for the Southbound API. Among them, there are traditional solutions like OpenFlow which is a binary Application level protocol structured to transport the instructions to be sent to the devices in its messages. NETCONF bases its operations on top of a simple Remote Procedure Call (RPC) layer. Configurations are encoded using an Extensible Markup Language (XML) and are serialized over secured transport channels. Instead, RESTCONF is based on HTTP. Consider the high diversity in the Southbound API solutions, the technological question "what is the best way to implement a Southbound API for SRv6?" is still open and different solutions have been proposed so far (for example see \cite{ovs_sr}). For this reason, we have considered and implemented four different variants of the SRv6 Southbound API (see Section~\ref{sec:propSouthAPI}), each one using a different transport mechanism and we have carried a performance evaluation in Section \ref{sec:experimental} to address the above question. 

\subsection{Comparison with OpenFlow based SDN solution}

The proposed SDN architecture for SRv6 enabled networks foresees the coexistence of distributed routing protocols with an SDN based approach. We can compare it with the more general OpenFlow based SDN approach which offers a great flexibility, enabling the classification of the packets through a "cross-layer" approach. To give an example OpenFlow allows to specify a set of matching conditions to influence the processing of the packets by considering packet headers at different protocol levels (MPLS, VLANs, Q-in-Q, Mac-in-Mac and so on). However, this flexibility may turn into high complexity and the risk of mis-configurations and routing errors should be properly taken into account (see \cite{vissicchio2013safe}). Another side effect, it is the lack of wide support of the aforementioned capabilities from most of vendors which can easily transform in a vendor lock-in. 

On the other hand, the SRv6 approach considers the classification based on Forwarding Equivalence Class or on the receiving interface at the ingress edge of an SRv6 domain and coexistence in the core with non-SRv6 traffic is based on the addition of the SRv6 header to the packets. This approach can effectively support important use cases as shown in \cite{segment-routing.net} and it is entirely based on IPv6 which represents the technology at the ground of the IP networks for the following years. Moreover, the SRv6 approach closely resembles the coexistence in the data-plane we currently have between IP and MPLS technologies which has proved to be a winning solution over the last twenty years.
\section{Southbound API for SRv6}
\label{sec:southboundAPI}

In this section, we review the state of the art of the SDN Southbound APIs, then we present our proposal (\ref{sec:propSouthAPI}) highlighting the most important implementation details. There are several types of Southbound APIs which are designed for different goals. Some of them focus on traffic management and rule enforcement while the others facilitate the process of configuration of network devices. In the following we briefly introduce the most famous ones.

OpenFlow (OF), developed by the Open Networking Foundation (ONF), is one of the most well-known Southbound interface and is considered as the first SDN standard for flow entries enforcement. Through this interface, the SDN Controller pushes down changes to the flow-table of switching/routing devices. This allows network administrators to partition traffic, control flows for optimal performance, and start testing new configurations and applications~\cite{mckeown2008openflow}. There are numbers of switch and router vendors that have announced their support of OF, including Cisco, Juniper, Big Switch Networks, Brocade, Arista, Extreme Networks, IBM, Dell, NoviFlow, HP, NEC, among others. It is notable that almost all controllers support OF. 

While OF is a flow entries enforcement API, Open vSwitch Database (OVSDB)~\cite{rfc7047} is a programmatic management protocol interface which is now being supported by network vendors, such as Cisco, Cumulus, Arista, and Dell. Cisco OpFlex~\cite{smith-opflex-03} is a mechanism to transfer abstract policy from a network controller to a set of smart devices capable of rendering abstract policy. The goal is to enable policies to be applied across physical and virtual switches/routers in a multi-vendor environment. Cisco One Platform Kit (OnePK) is another Southbound protocol used in Cisco devices to develop, automate, and rapidly create a service. However, it is not supported any more by Cisco~\cite{CiscoOnePK}.

Path Computation Element Communication Protocol (PCEP)~\cite{rfc5440} is another Southbound API for creation of Label Switched Path (LSP) in MPLS networks. More in details, it is a special set of rules that allows a Path Computation Client (PCC) to request path computations from Path Computation Elements (PCEs). Most of the available SDN controllers (commercial and non) support PCEP. The Network Configuration Protocol (NETCONF) is a network management protocol developed and standardized by the IETF for accessing data defined in YANG~\cite{rfc6241}. It provides mechanisms to install, manipulate, and delete the configuration of network devices on top of a simple RPC layer. It leverages SSH as transport mechanism and uses an Extensible Markup Language (XML) to communicate with the routing devices to install and make configuration changes. Instead RESTCONF \cite{rfc8040} is a REST like protocol running over HTTP and defines the mapping of a YANG specification to a RESTful interface.

REST/HTTP interfaces have been historically used on the northbound side of a SDN controller, recently it is possible to find a number of devices exposing a RESTful interface towards the controllers \cite{cienawaveserver}. gRPC \cite{grpc} is a modern RPC framework initially developed by Google and then run by an active community of developers. It uses the modern HTTP/2 for transport, Protocol Buffers \cite{protobuf} as the interface description language, and provides out-of-box interesting functionalities like authentication, bidirectional streaming, flow control and many others. Then, there are a number of solutions leveraging Thrift \cite{thrift} as RPC mechanism to implement the Southbound APIs. For example, FBOSS from Facebook \cite{fboss} provides Thrift APIs to allow external routing processes (BGP or SDN Controller) to get their routes programmed into the hardware forwarding tables.

In the SDN research field, usually the focus has been on proposing new controller architectures and TE applications; an exhaustive survey of research activities on SDN is provided in \cite{kreutz2015software}. Some novel works focus on integrating SDN with Segment Routing see for example~\cite{salsano2016pmsr,pang2017sdn,sheu2017scalable,davoli2015traffic}. For a comprehensive survey of research activities, standardization efforts and implementation results on Segment Routing see~\cite{srsurvey}. To the best of our knowledge, there are no research efforts among the analyzed works proposing a Southbound API for Segment Routing.

\subsection{Implemented Southbound APIs}
\label{sec:propSouthAPI}

The process of configuring the SRv6 rules in the routing devices can be decomposed in two aspects: the communication protocol and the local configuration of the rules. The SDN controller uses the communication protocol to send the requests to the SRv6 Manager running on the node. Based on the received messages, the SRv6 Manager takes care of the SRv6 commands needed on the routing device to configure the rules.

We have focused on the implementation of the SRv6 SDN Southbound API considering four different variants: i) gRPC, using the Protocol Buffers Interface Definition Language (IDL) and HTTP/2 as transport mechanism; ii) RESTful approach over HTTP/1.1, leveraging JSON notation for describing data; iii) NETCONF, carrying the configuration in XML format; iv) using a wrapper module around the CLI (Command Line Interface) that performs the commands remotely over SSH. Both the gRPC and REST API are based on HTTP and can additional provide security through SSL/TLS layer, while the NETCONF and CLI API use SSH as transport mechanism. From an architectural point of view, all the realized APIs result to be similar, since both have a client running inside the SDN controller and the server process running in the SRV6 network node. As regards the CLI/SSH implementation, we spawn a new server inside the node and we did not reuse the SSH daemon already running inside the device. This has been done in order to have more control of the operations. The CLI/SSH Southbound API results to be more limited compared to the other implementations, because it can only transport commands which can be run inside the node. On the other hand, the gRPC, REST and NETCONF implementations are more flexible and more expressive since they provide the means to realize a more structured API with parameters and fields and not just a big string carrying on the entire command. For all the four API variants we have implemented a python library/wrapper that can be included in a python SDN controller implementation.

\begin{figure}
    \centering
    \includegraphics[width=0.485\textwidth]{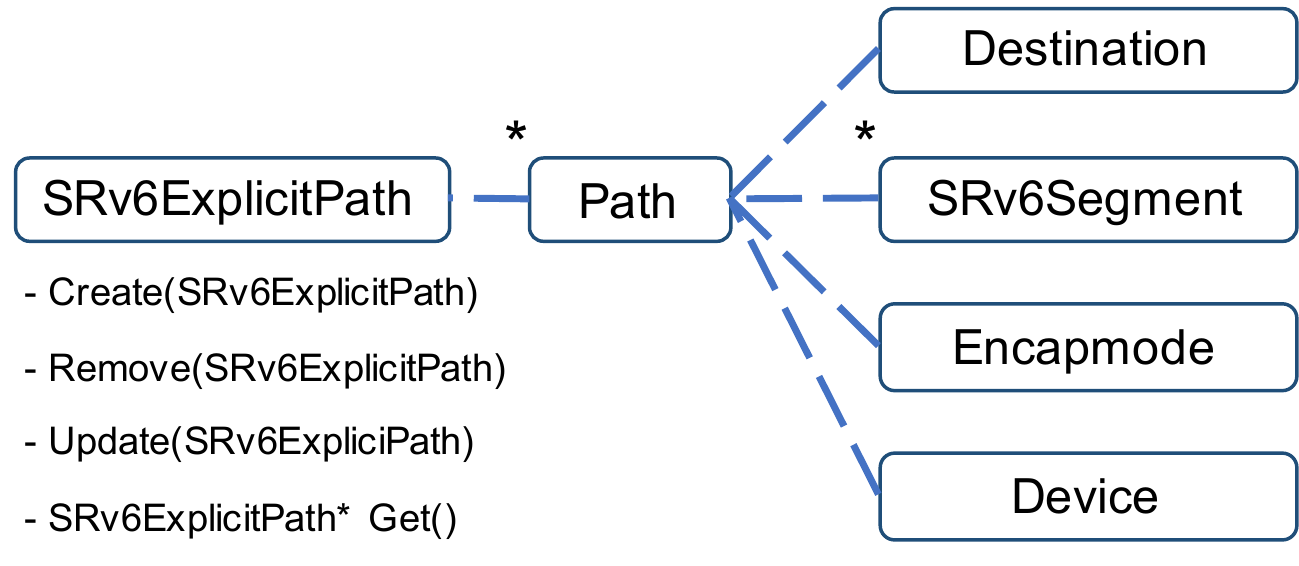}
    \caption{Southbound API data model and interfaces}
    \label{fig:sb_dm}
    \vspace{-3ex}
\end{figure}

Taking as reference the interface of \textit{iproute2}, we have defined the interfaces and the parameters and the fields of the messages to be transported over the connection which are then serialized over the wire. Figure \ref{fig:sb_dm} shows the representation of the data-model we have used together with the definition of the supported interfaces. More specifically, the RESTful implementation implements a JSON-RPC mechanism and basically translates the data-model defined in Figure \ref{fig:sb_dm} in a JSON object and send this data over an HTTP \textit{POST} to the server. The different services to be called on the server are accessible using as base path of the URL \textit{srv6-explicit-path} and adding the parameter \textit{operation} in the query string of the URL, which can take as value: create, remove, update and get.

NETCONF protocol has been designed to modify the configuration of the devices and offers an already defined protocol, in this case the implementation of the API has been straightforward since we were constrained by the framework defined by the protocol. In particular the client does an \textit{<edit-config/>} RPC on the device and conveys in the message the data-model; \textit{<srv6-explicit-path/>} is sent as root element and \textit{operation} is a parameter of it. As regards the \textit{get} operation we leverage the \textit{<get-config/>} RPC to dump the running configuration of the device. As regards gRPC, it offers a better way to design and structure a protocol thanks to the use of the Protocol Buffers which drive also the serialization of the data. The serialized data are then sent as binary over the TCP connection. With respect to plain HTTP, gRPC requires a further step since all the interfaces and the messages need to be defined creating a \textit{proto} file \cite{protobuf}. In particular, we have defined a \textit{Service} offering 4 RPCs which allows to add, remove, get and change SRv6 configuration in a SRv6 Linux based node. For the add, remove and change RPCs we have defined a \textit{Request} message which basically carries a variable number of \textit{Path} as showed in Figure \ref{fig:sb_dm} and a \textit{Reply} to report the status of the operation. As regards the get operation, there are no input parameters and the \textit{Reply} message returns the SRv6 routes installed in the device. In the future, the gRPC implementation of the SRv6 Southbound API can be used as base and easily improved thanks the ecosystem provided by the gRPC project. In our current roadmap, we plan to focus on the gRPC based API, while the other variants of the API have been developed mainly for comparison. In fact, gRPC for SDN seems to be gaining momentum as witnessed by the recently announced Stratum project \cite{stratum} run by ONF which aims at ``enabling the era of next generation SDN''. Our proposed SRv6 SDN framework is available at \cite{srv6-sdn}. 

It is worth noting that we did not consider an OpenFlow based design for the SRv6 Southbound API like in \cite{ovs_sr}. From the technical standpoint it is much more complex to design and implement an OpenFlow based solution with respect to the options that we have considered, making it a less preferred solution. The theoretical advantage of OpenFlow lies in its potential multi-vendor and multi-technology support, but this require a standardization procedure which is long and complex (and out of the reach of academy institutions). It also worth mentioning another multi-vendor solutions, namely OpenConfig \cite{OpenConfig} which could address most of the needs covered by OpenFlow. 
\section{Intent based emulation of SRv6 networks}
\label{sec:emulation}

We have released a set of tools that simplify the emulation of a SRv6 enabled networks. The management tools are a collection of projects meant to support SRv6 experiments both over physical deployments, Mininet and virtual testbeds. As regards the latter, the tools are enough generic to support different type of providers like Amazon, Azure and others; since the minimal requirement is virtual testbeds which offer VMs as resources and connectivity between them. These tools offer means to design, control and measure several aspects of an experiment, they include: i) a web GUI to design the topology to be emulated, ii) deployment scripts to configure the nodes, iii) the possibility to interact with the emulated nodes through the web GUI; iv) the possibility to run experiments over the emulated topology. The overall emulation framework is described at \cite{rose}, in which the instruction to download the tools and setup of the emulation environment are provided. 

\begin{figure}
    \centering
    \includegraphics[trim={6.5cm 0.5cm 8cm 0cm},clip,width=0.43\textwidth]{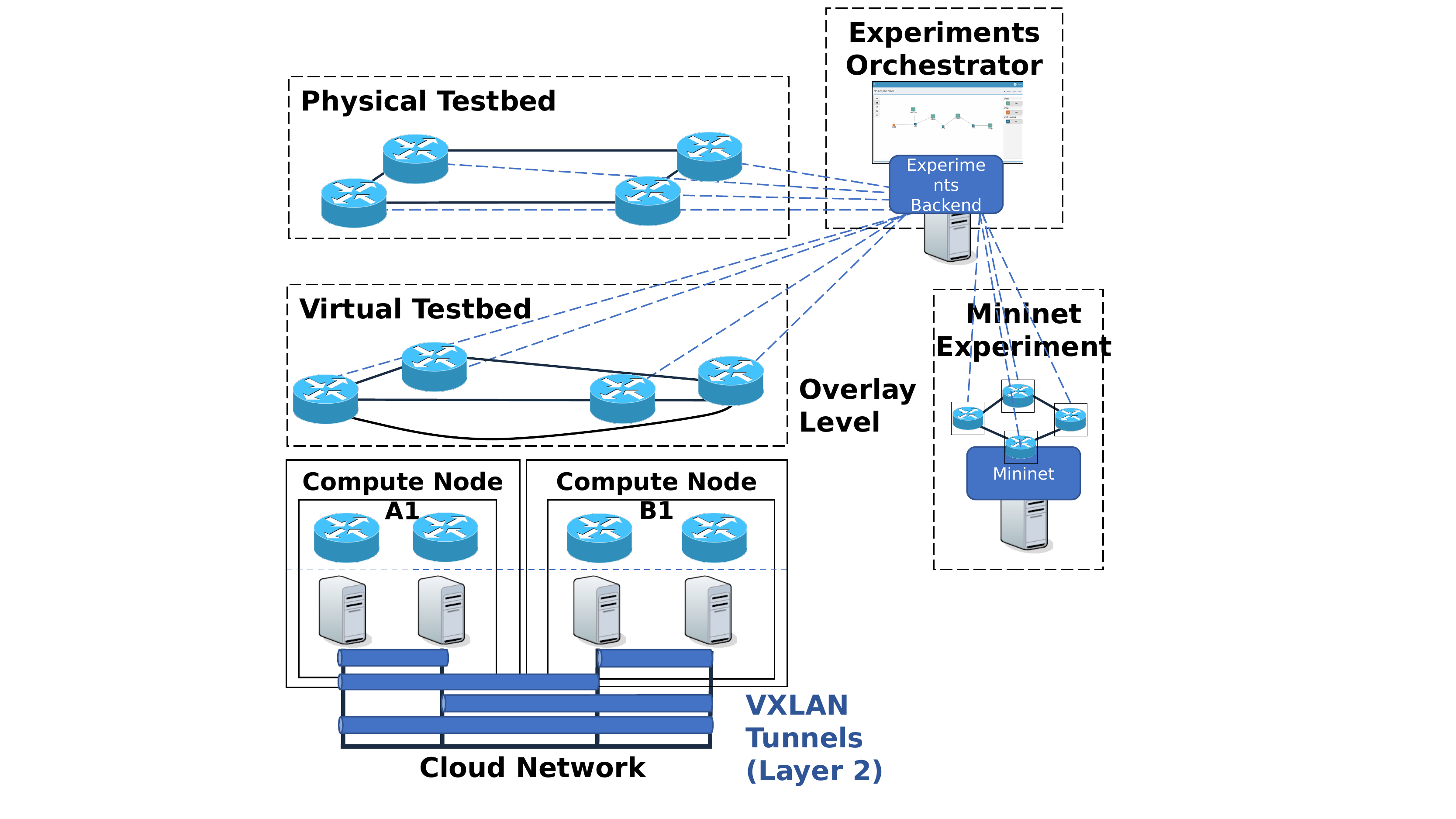}
    \caption{Intent based emulation tools}
    \label{fig:srv6_emul_tools}
    \vspace{-3ex}
\end{figure}

Figure \ref{fig:srv6_emul_tools} shows the management tools in action and the supported scenarios; in the bottom part of the figure is shown the deployment of an overlay experiment over a virtual testbed managed by a Cloud infrastructure. It is worth noting that we can use the same software to run Mininet emulations or deploy an experiment over a physical deployment made of Linux boxes. For example using this framework, we have deployed SRv6 experiments \cite{tiesr} over SoftFire testbed \cite{softfire} realizing Service Function Chaining scenarios which included SRv6/OSPFv3 routers controlled by a SDN controller.

The emulation consists in a network composed of 4 types of nodes: SRv6 enabled routers, terminal nodes, VNF nodes, and a SDN controller. Each terminal and VNF node are connected with one router (we applied a well-established principle of having a single ``default via'' for them). The controller can be connected with one or more routers. The routers can be arbitrarily interconnected with other routers. The emulation framework provides the automatic definition of the IPv6 addressing plan and the proper configuration and deployment of dynamic routing among the SRv6 enabled routers (using OSPFv3). 

As mentioned before, two types of emulations are supported: i) Mininet emulation; ii) distributed emulation with Virtual Machines (VMs) or physical nodes. In the Mininet emulation case, all the 4 node types are deployed as Mininet containers running inside a single Linux host. In the distributed emulation case, we assume that a set of Linux hosts (VMs or physical machines) is available to run the emulated nodes. In particular, on each box we deploy an SRv6 based router, the set of terminals and VNFs that are connected to the router. The terminals and VNF nodes can be deployed as Linux network namespaces or as Linux containers (they are technically containers running inside the SRv6 routers). The links among the SRv6 routers in the emulated topology are realized using VXLAN tunnels (the blue pipes in Figure \ref{fig:srv6_emul_tools}), which are automatically setup and configured by the management tools leveraging the underlay connectivity provided by the IAAS infrastructure. In the case of physical testbeds the management scripts leverage directly the physical connection of the machines. 

All these steps, from the generation of the configuration up to its ``implementation'' have been completely automated. A management host (experiments orchestrator in Figure \ref{fig:srv6_emul_tools}) coordinates the overall process. In particular, it allows users to express their intents regarding the emulation using a topology GUI and codifies these desires through a graph. Then, this representation is given as input to the emulation engine which defines, for each machine participating in the experiment, its role and its configuration together with IPv6 addressing plan. Moreover, in the case of a deployment over a virtual testbed it takes over the task of properly defining the VXLAN tunnels according to the links of the emulated topology. For each machine, it creates a configuration file which is pushed on the proper VM and then, using local management scripts (deployed during the provisioning), the configuration is made effective. In this way, the experimenter is relieved from a huge configuration burden and can focus only on the SRv6 related aspects of the experiment. To further simplify the setup, we also provide a VM image in which all the developed components have been pre-installed \cite{rose}.

\section{Experimental evaluation}
\label{sec:experimental}

In this section, we present an evaluation of some performance aspects of our architecture. At first, we analyze the local configuration performances in our Linux based SRv6 device (Section \ref{sec:lre}), then we provide a comparison of the different implementations of the Southbound APIs (Section \ref{sec:comparison}) and finally we evaluate the effects of the dynamic reconfiguration of SRv6 policies on active flows in an emulated network-wide scenario (Section \ref{sec:dynamic}). 
For the local configuration performance and for the comparison of the different Southbound APIs, the considered metrics are the configuration execution time (or response time for remote configurations) and the CPU and memory utilization in the SRv6 device. For the effects of the dynamic reconfiguration we considered the packet loss and compared the received traffic profile with the traffic profile configured by the controller.
In order to run the considered experiments we developed a set of measurement tools. They include two stub servers for enforcing the configuration commands in the SRv6 device, respectively using \textit{pyroute2} or \textit{shell} commands. These Linux applications represent two reference implementations of the SRv6 Manager and include timestamp recordings for the main steps of the experiments. We have developed different client/server applications for the analysis of the different communication protocols. For the evaluation of CPU and memory usage, we developed simple scripts to record the system CPU usage along with the total memory usages inside our SRv6 nodes. Finally, for the evaluation of the effect of the dynamic reconfiguration of SRv6 policies we have implemented a simple controller application that enforces the mapping of a flow into different SRv6 tunnels with a predefined timing. All implemented measurement tools are open source and available on \cite{srv6-sdn}.

\subsection{Local Rule Enforcement}
\label{sec:lre}

The purpose of this experiment is to characterize the performance of the execution of the configuration commands on the Linux SRv6 device in isolation. In other words, there are no control messages received from the SDN controller and the whole operations are taking place locally. For this experiment we have used a laptop equipped with an Intel Core Duo 2.4 Ghz dual core and 4GB of RAM. The machine uses a Linux distribution as OS where we installed the kernel 4.15.
In particular, we compare two python based variants of the back-end for the SRv6 Manager, one uses the \textit{pyroute2} library to interact with Linux kernel networking, another one (called \textit{shell} hereafter) enforces the local configuration through the \textit{ip route} command executed in a shell. 

The execution time of a single command can be very small, therefore we execute a number $N_i$ of identical commands and measure the total execution time. We repeat the experiment $M_i$ times and estimate the average, the Coefficient of Variation (CV) and the 95\% Confidence Interval ($CI_{95}$). As described in section \ref{sec:SRv6node}, the \textit{pyroute2} variant opens a \textit{netlink} socket with the Linux kernel to request the configuration operations. A single \textit{netlink} connection is opened at the beginning of the experiment and it is used to send the $N_i$ configuration commands.

Table~\ref{tab:PyrouteVSShell} reports the mean $\mu$, CV and $CI_{95}$ of the execution time for both \textit{add} and \textit{delete} operations and compares the performance of \textit{pyroute2} and \textit{shell} approaches. We enforced $N_i=100$ operations and repeated each experiment $M_i=20$ times.

\begin{table}[!ht]
    \centering
    \begin{tabular}{|l|c|c|c|c|c|c|}\hline
        \multicolumn{7}{|c|}{Execution Time of 100 operations (s)}\\\hline
        & \multicolumn{3}{|c|}{Add} & \multicolumn{3}{|c|}{Delete} \\\hline
        & $\mu$ & CV & $CI_{95}$ & $\mu$ & CV & $CI_{95}$ \\\hline
        pyroute2 & 0.06 & 6.9$\%$ & 3.08$\%$ & 0.05 & 6.7$\%$ & 3$\%$ \\\hline
        shell & 0.28 & 2.9$\%$ & 1.29$\%$ & 0.27 & 2.0$\%$ & 0.89$\%$ \\\hline
    \end{tabular}
    \caption{pyroute2 vs. shell (average of 20 runs)}
    \label{tab:PyrouteVSShell}
    \vspace{-2ex}
\end{table}

As can be seen, the execution time of \textit{delete} operation is lower than \textit{add} operation in both approaches. Comparing \textit{shell} and \textit{pyroute2}, the response time of \textit{pyroute2} is lower than \textit{shell}. This result is expected because the \textit{shell} approach executes most of the operations via external commands while the \textit{pyroute2} version runs the configurations in the same user process opening a \textit{netlink} socket to talk with the kernel. 

From these experiments we can estimate an upper bound of the maximum configuration rate of the SRv6 device. Considering the \textit{pyroute2} variant, the results reported in Table~\ref{tab:PyrouteVSShell} correspond to around 1700 add operations per second or more than 2000 delete operations per second. The maximum configuration rate for the \textit{shell} approach is substantially lower, i.e. in the range of 350 operation per second (for add operations). Of course these results are dependent of the hardware capability (in particular CPU power) of the SRv6 device under analysis. Anyway, even by scaling down 5 or 10 times the processing capabilities of the device (i.e. considering low end devices) in our opinion the configuration rate should remain acceptable for the considered devices, in the order of 150-300  configuration operations per second.

\begin{figure}[!htbp]
    \centering
    \begin{subfigure}{0.49\columnwidth}
	    \centering
        \includegraphics[width=1\columnwidth]{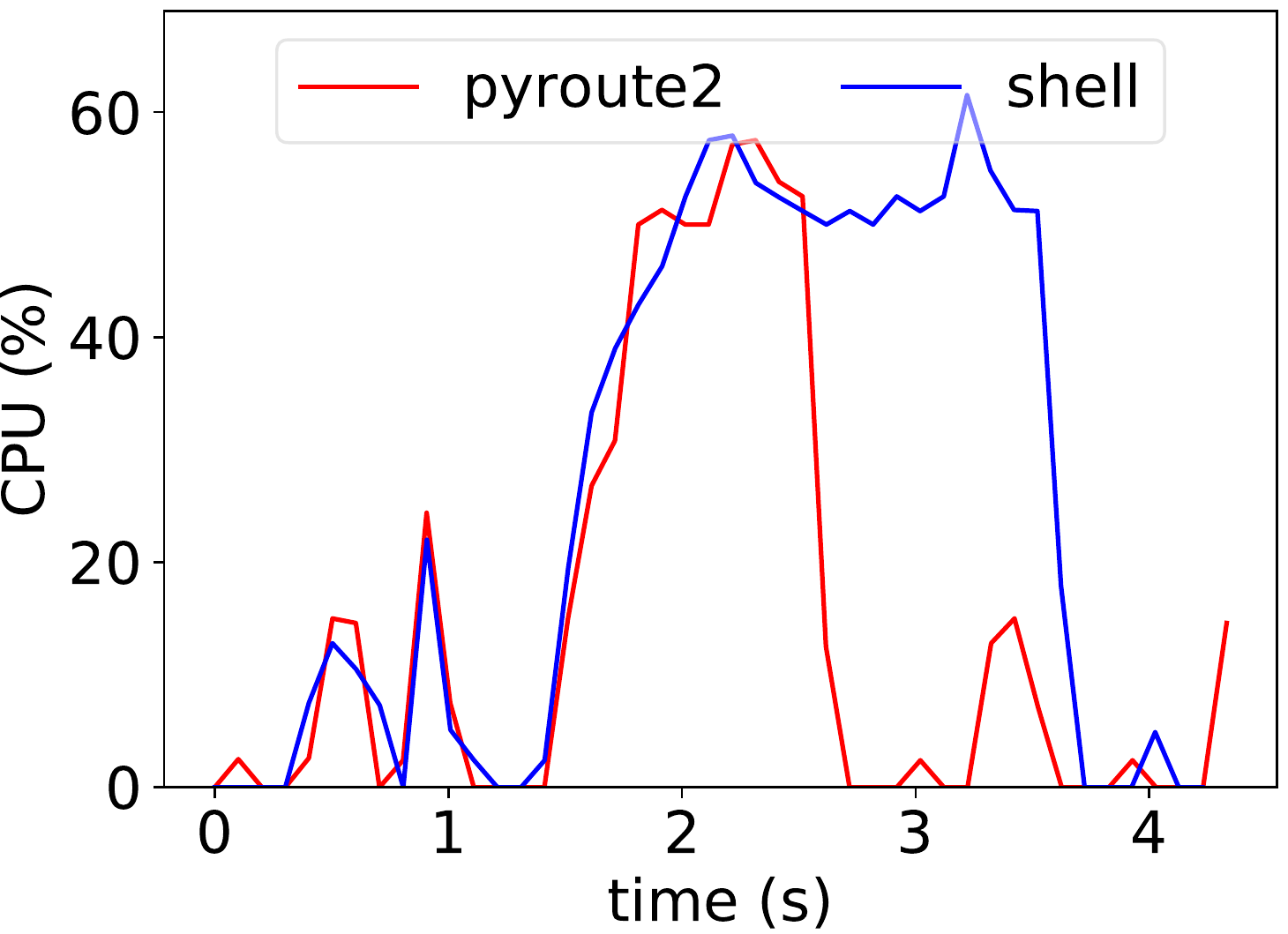}
        \caption{CPU usage} 
        \label{fig:cpu-pyroute-shell}
    \end{subfigure}
    \begin{subfigure}{0.49\columnwidth}
	    \centering
        \includegraphics[width=\columnwidth]{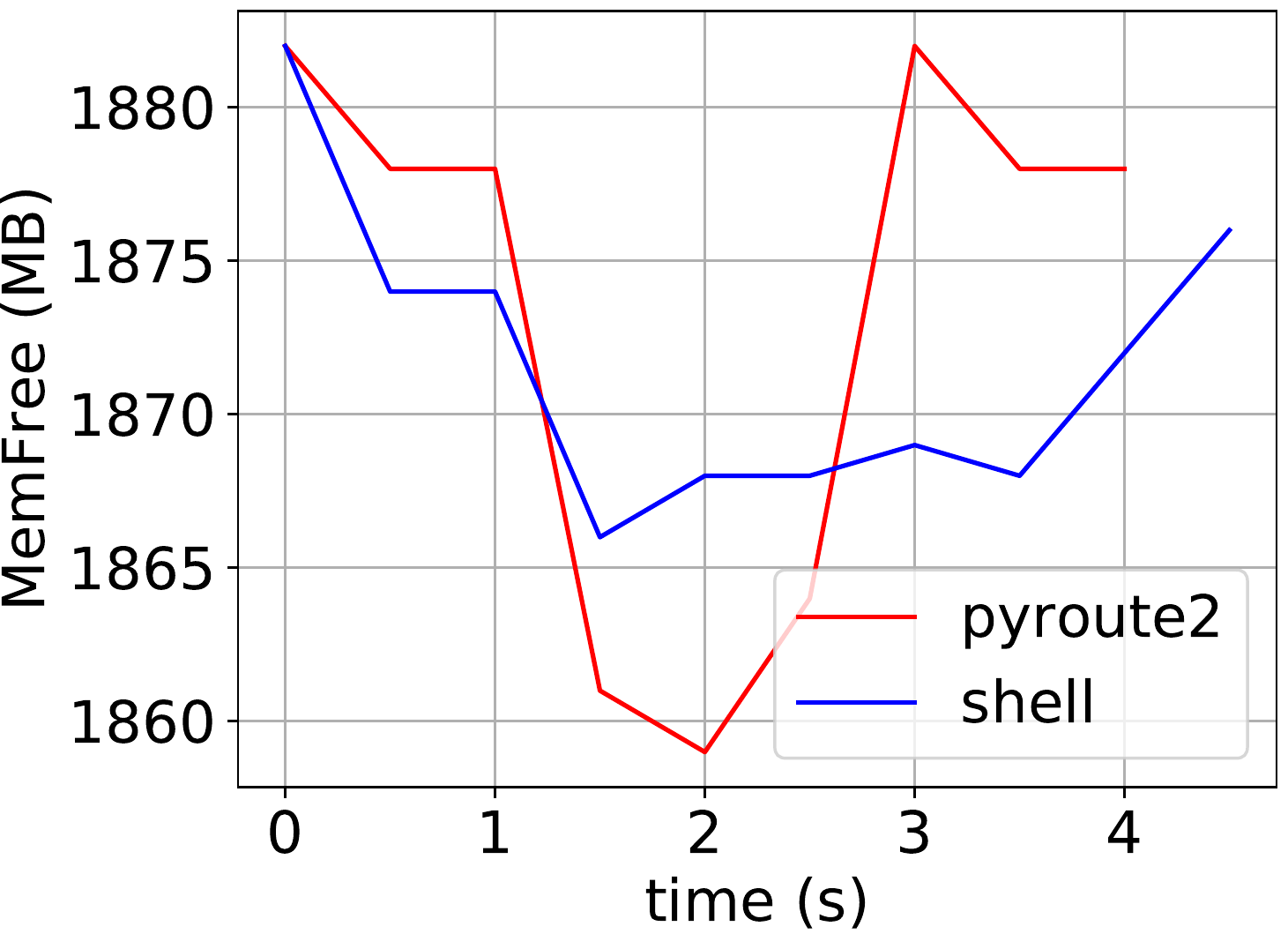}
        \caption{Free memory}
        \label{fig:memory-pyroute-shell}
    \end{subfigure}
    \caption{pyroute2 vs shell (1000 add enforcements - 1 run)}
    \label{fig:shellvspyroute}
    \vspace{-1ex}
\end{figure}

\begin{figure*}[!htbp]
    \centering
    \begin{subfigure}{0.62\columnwidth}
	    \centering
        \includegraphics[width=1\columnwidth]{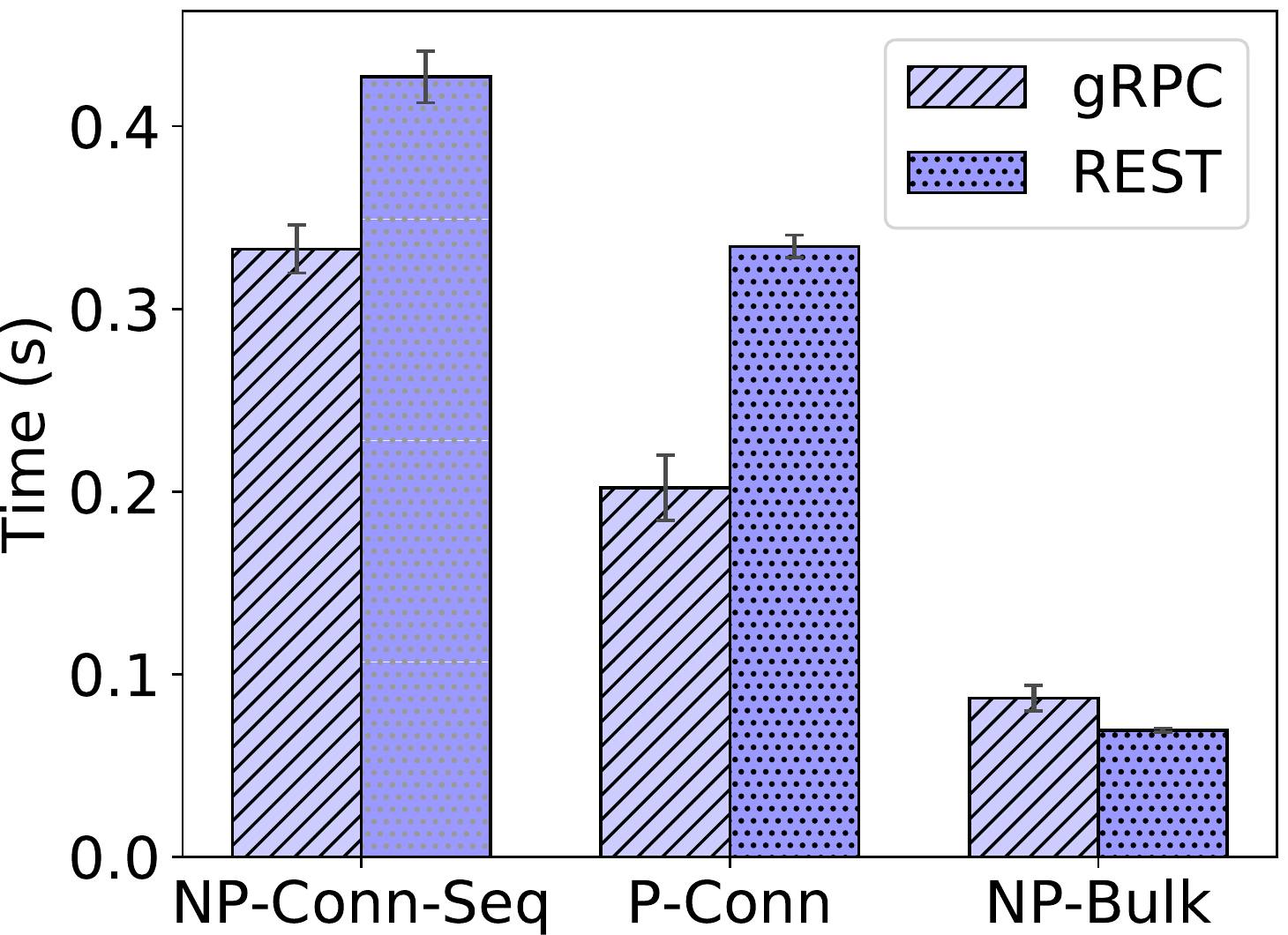}
        \caption{Insecure connections}
        \label{fig:full-insecure}
    \end{subfigure}
    \begin{subfigure}{0.62\columnwidth}
	    \centering
        \includegraphics[width=1\columnwidth]{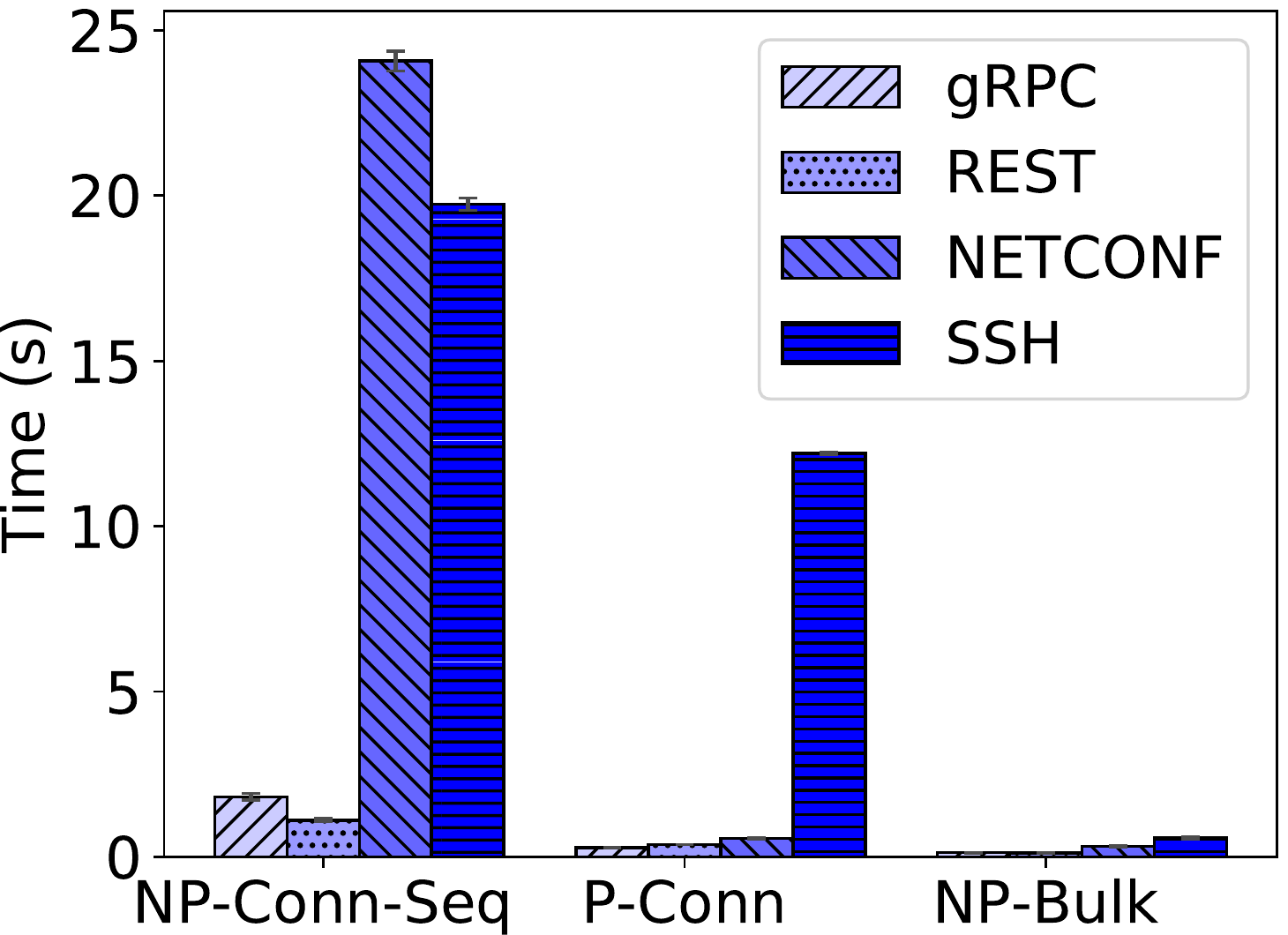}
        \caption{Secure connections}
        \label{fig:full-secure}
    \end{subfigure}
    \begin{subfigure}{0.62\columnwidth}
	    \centering
        \includegraphics[width=1\columnwidth]{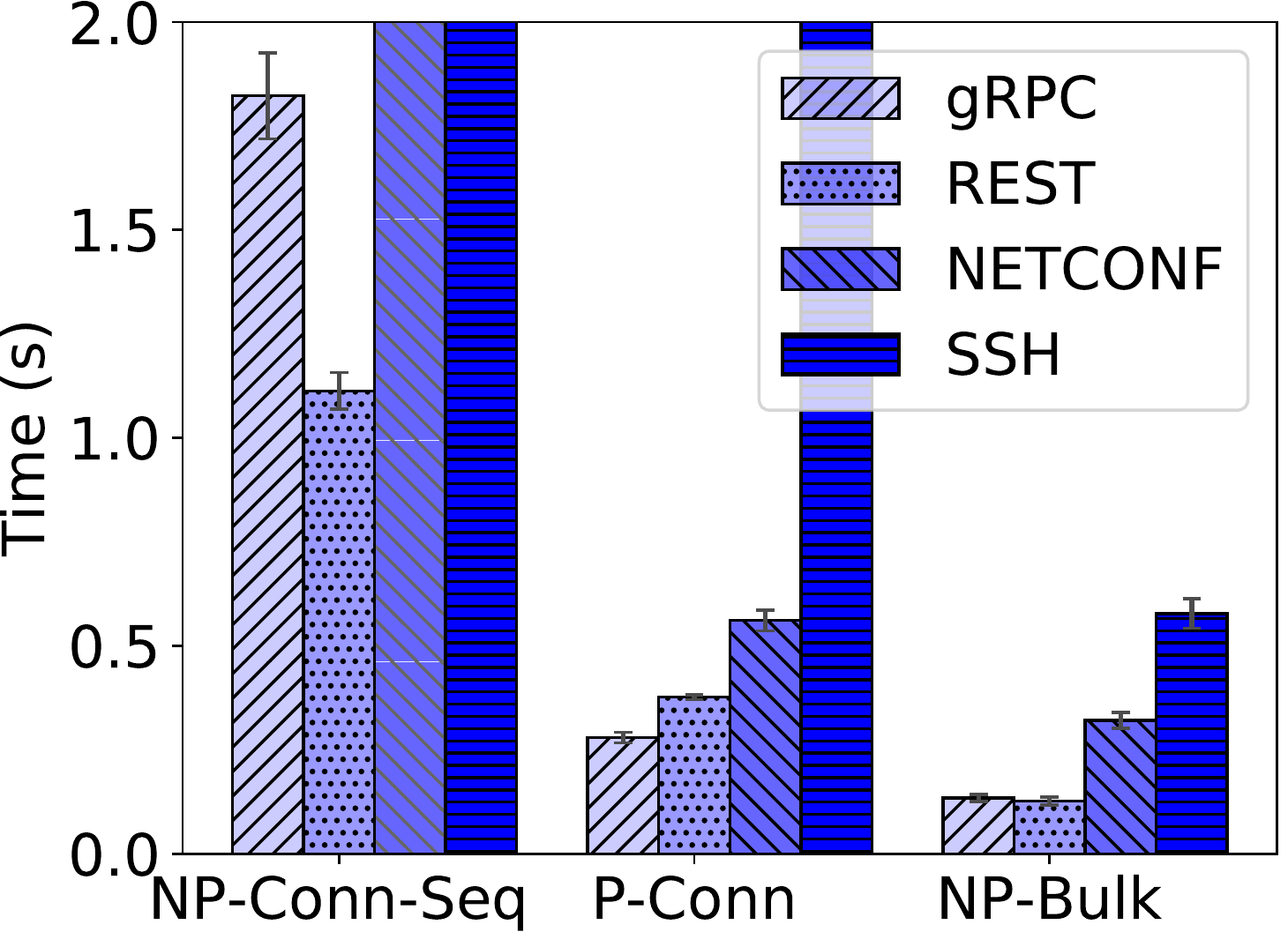}
        \caption{Secure connections (zoomed in)}
        \label{fig:full-secure-zoom}
    \end{subfigure}
    \caption{Full Config - Response time}
    \label{fig:full-response-time}
    \vspace{-3ex}
\end{figure*}

We have run another type of experiments to analyze the CPU and memory utilization of the \textit{pyroute2} and \textit{shell} based approach. For the CPU utilization, our main goal was to verify that the overall system performance is CPU-limited in the experiments reported above. In this experiment, reported in Figure~\ref{fig:shellvspyroute}(a), we execute $N_i=1000$ add operations to have a longer observation interval (around 1 second for \textit{pyroute2} and 2 seconds for \textit{shell}). The total system CPU load ($\%$) over time is reported in the figure. As the system has 2 cores, the load of $50\%$ correspond to the full utilization of 1 core. The configuration experiment starts at time 1.2 s in Figure~\ref{fig:shellvspyroute}(a), when the CPU load starts to increase steeply for both \textit{pyroute2} and \textit{shell}. The CPU load remains at around $50\%$ for one second for \textit{pyroute2} and two seconds for \textit{shell} (the oscillation above $50\%$ are due to other processes on the other core).

The results shown in Figure~\ref{fig:shellvspyroute}(a) confirm that the overall CPU work for $N_i=1000$ add operations for \textit{pyroute2} is the half of the work for \textit{shell}, as the latter has a double duration with the same CPU load. Finally, in Figure~\ref{fig:shellvspyroute}(b) we analyze the memory utilization. The results show that the memory (RAM) utilization is relatively low and it is not a concern in typical device configurations. In particular, Figure~\ref{fig:shellvspyroute}(b) shows the overall free memory in the device sampled every 500 ms while running the two experiments with $N_i=1000$ add operations. The memory used of the \textit{pyroute2} approach is in the order of 15 MBytes, while the memory used by the \textit{shell} approach is 10 MBytes. This can be explained by the nature of the execution of \textit{shell} approach which leverages $N_i$ separate configuration commands which end after the enforcement of the configuration. However this difference is practically negligible.

\subsection{Comparison of the Southbound APIs}
\label{sec:comparison}

In this section we analyze some performance aspects of the four implementations of SRv6 Southbound API (gRPC, REST, NETCONF and SSH/CLI). We analyze different variants for the four implementations. In particular we consider three different interaction modes between the SDN controller and the SRv6 device: i) persistent connection (`P-Conn') in which several requests are sent reusing one single TCP connection, ii) non-persistent connections (`NP-Conn-Seq') in which the configuration requests are sent sequentially using a separate TCP connection for each request, and iii) bulk requests in which a number of configuration requests is sent on one single message (`NP-Bulk'). We note that the most common interaction mode when there is the need of sending several configuration messages over time should be the first one (`P-Conn'), in which the SDN controller establishes a TCP connection with the SRv6 device and reuses it for all the messages. When the SDN controller needs to interact sporadically with the devices, the other two interaction modes could make sense.

We consider \textit{insecure} and \textit{secure} connections. In particular for the \textit{insecure} connection scenario we consider only the gRPC and REST implementations, for which we can easily enable and disable the security mechanism (authentication and encryption). The NETCONF and SSH/CLI implementations are both based on SSH that includes security by default, so they only belong to the \textit{secure} scenario which allows comparing all the 4 implementations.

We analyze the Southbound API implementations in two ways. First, we include the execution of the configuration operations on the SRv6 device (\textit{Full Config} experiments). Then, in order to focus on the performance of the communication part we exclude the execution of the configuration operations in the SRv6 device (\textit{Communication Only} experiments). For the \textit{Full Config} experiments we use the \textit{pyroute2} approach for the interaction with the kernel in the gRPC, REST and NETCONF implementations, while in case of the SSH/CLI the commands are executed in a \textit{shell}. 

We have first considered the experiments in ideal conditions in which the SDN controller and the SRv6 device are physically close each other and connected over a LAN with negligible packet loss, then we have considered a scenario with network impairments by synthetically adding a one way delay of 75 ms between the SDN controller and the SRv6 device (corresponding to 150 ms of Round Trip delay) and different packet loss ratios ($0$, $0.5$, $1$, and $2\%$ on each direction).  We refer to this last scenario as \textit{NLD} (Network Loss and Delay). The topology of the experiments comprises two identical laptops equipped with an Intel Core Duo 2.4 Ghz dual core and 4GB of RAM. A recent Linux kernel (4.15) has been installed on the laptops. The laptops are connected with a point-to-point cable at 1 Gb/s. The SDN controller is installed on one laptop and the other laptop is acting as SRv6 device. The results reported hereafter are the average of $M_i$ runs.

\begin{figure*}[!htbp]
    \centering
    \begin{subfigure}{0.49\columnwidth}
	    \centering
        \includegraphics[width=1\columnwidth]{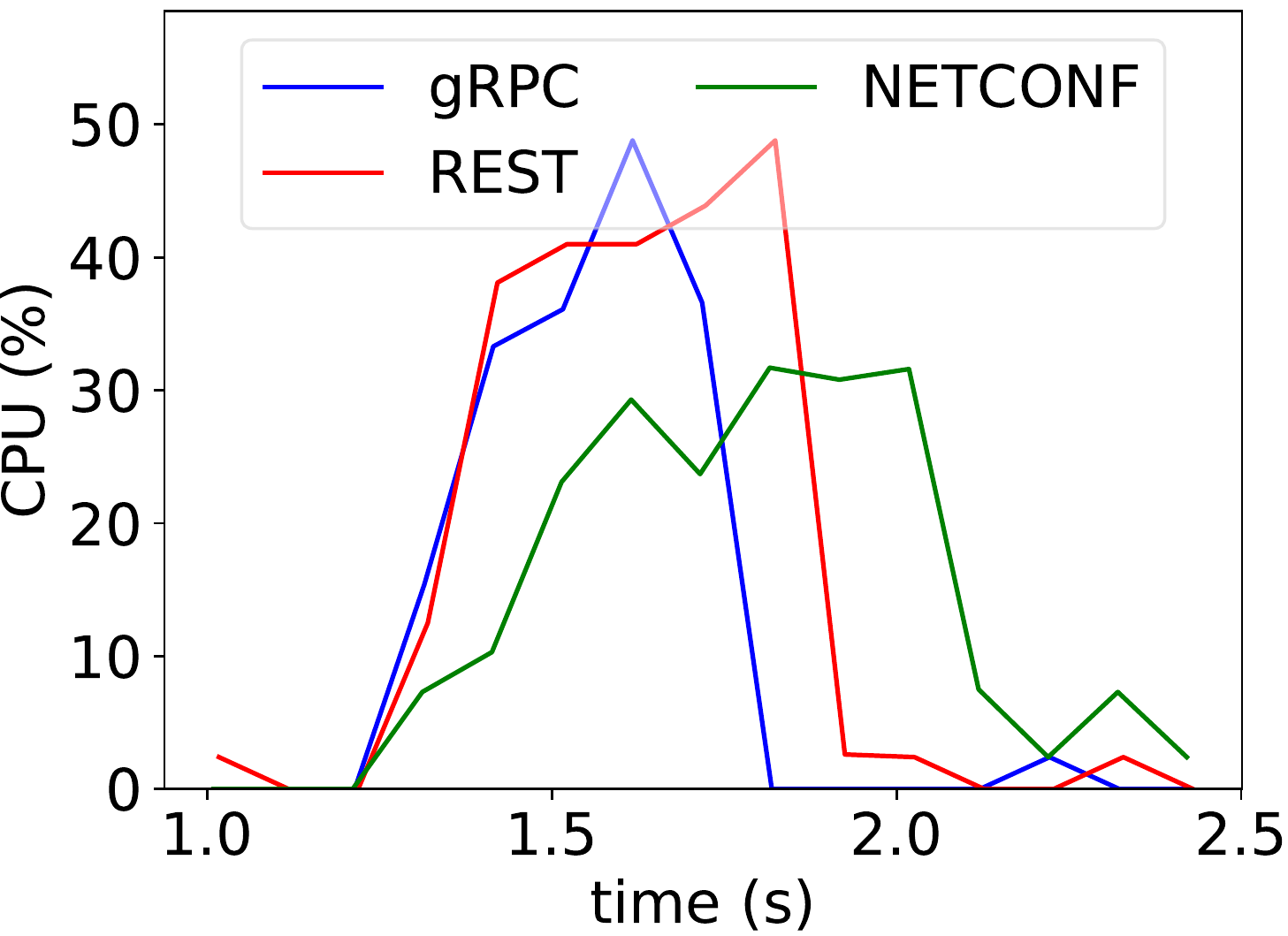}
        \caption{gRPC vs REST vs NETCONF}
        \label{fig:cpu-full-pc-grpcrestnetconf}
    \end{subfigure}
    \begin{subfigure}{0.49\columnwidth}
	    \centering
        \includegraphics[width=1\columnwidth]{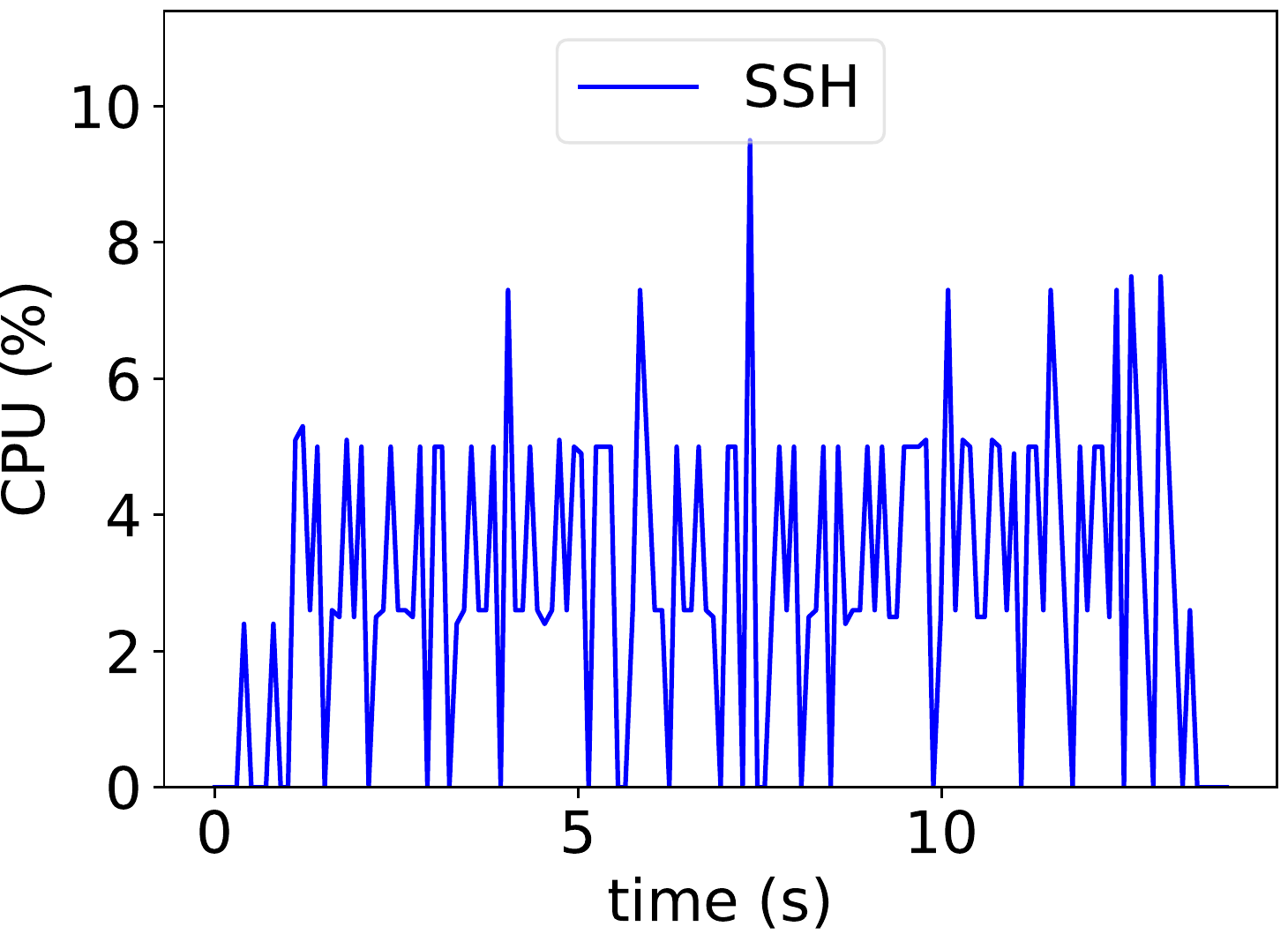}
        \caption{SSH}
        \label{fig:cpu-full-pc-ssh}
    \end{subfigure}
    \begin{subfigure}{0.49\columnwidth}
	    \centering
        \includegraphics[width=\columnwidth]{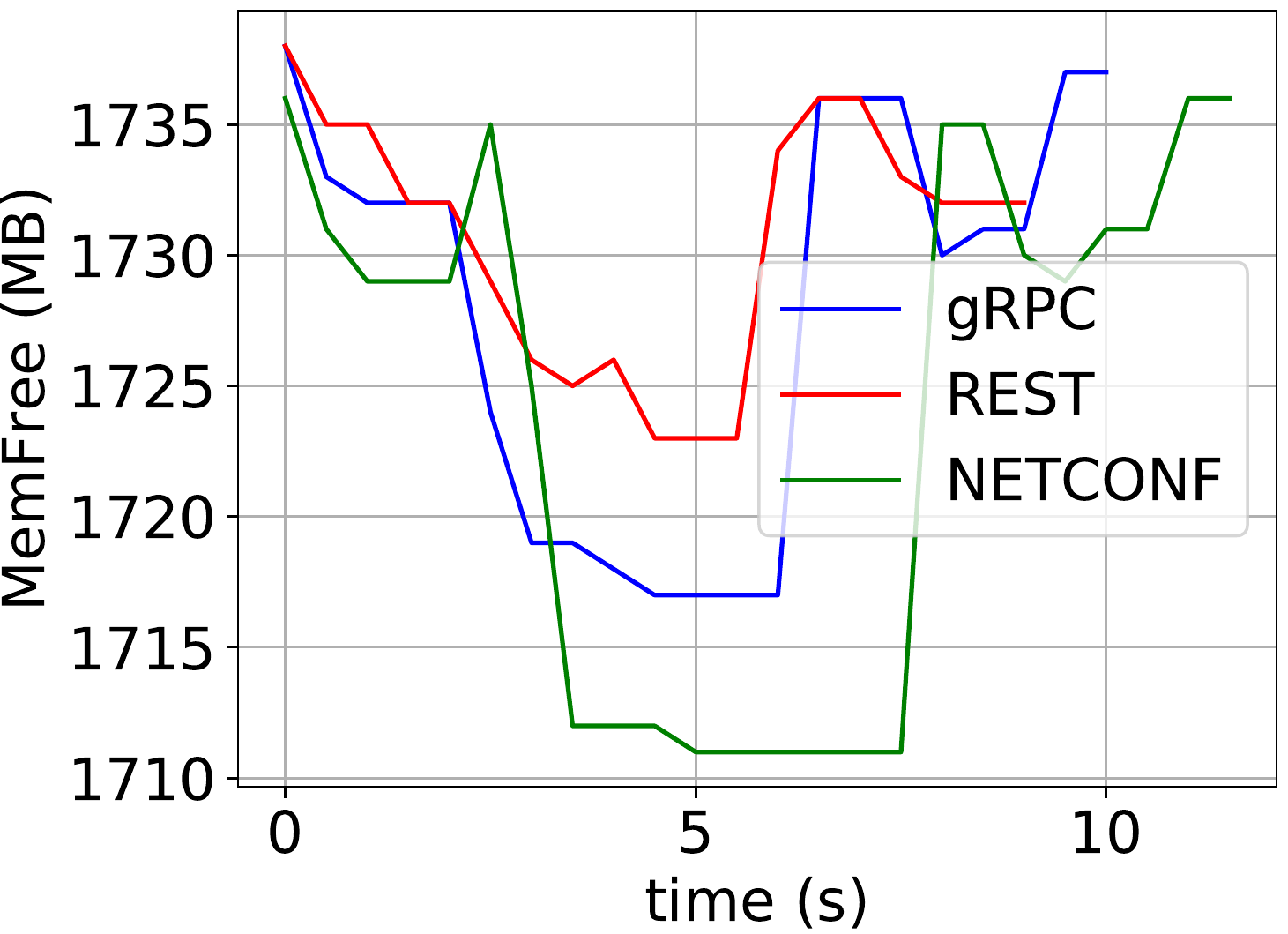}
        \caption{gRPC vs REST vs NETCONF}
        \label{fig:memory-full-pc-grpcrestnetconf}
    \end{subfigure}
    \begin{subfigure}{0.49\columnwidth}
	    \centering
        \includegraphics[width=\columnwidth]{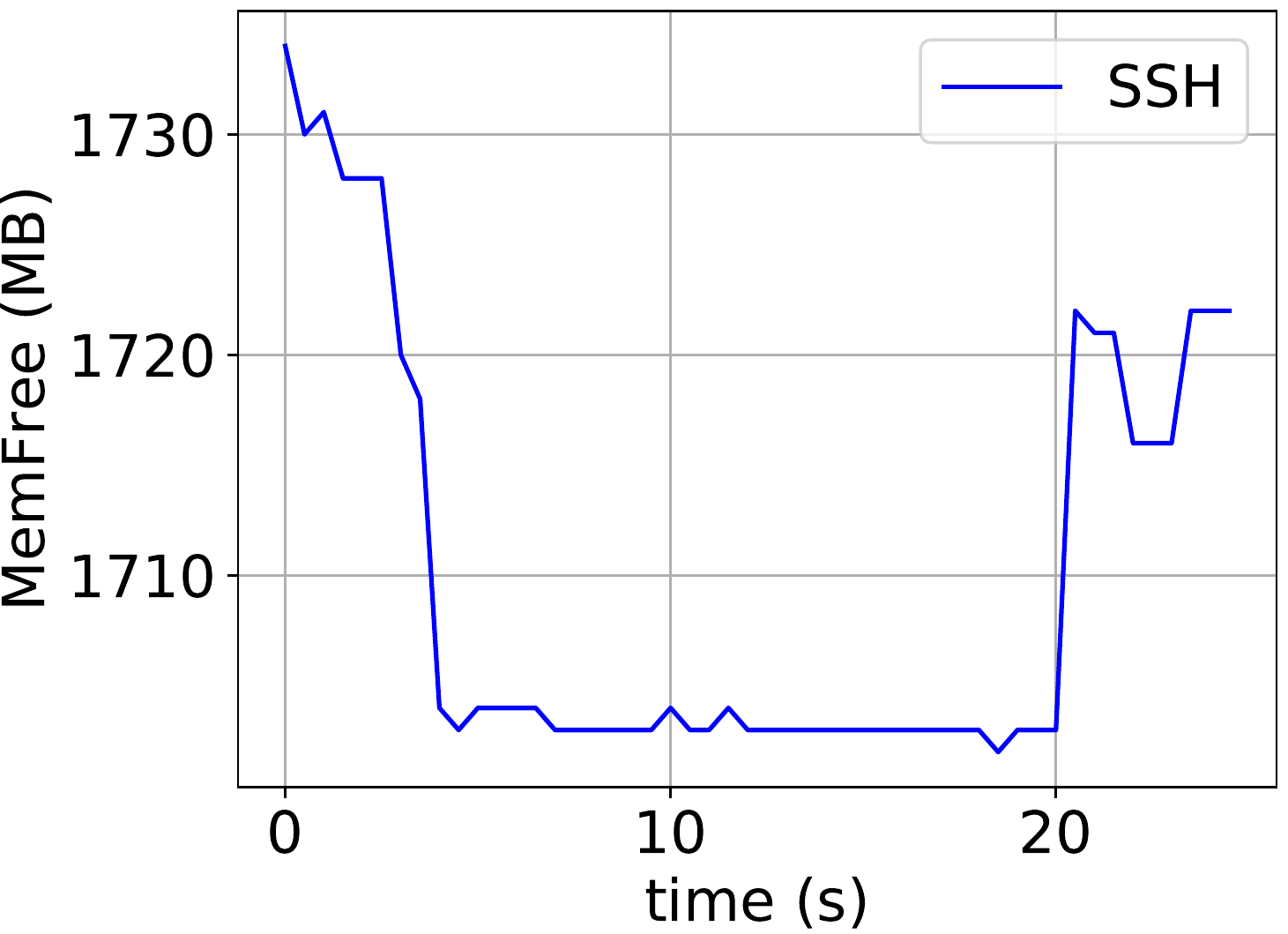}
        \caption{SSH}
        \label{fig:memory-full-pc-ssh}
    \end{subfigure}
    \caption{CPU and memory usage, secure Permanent connection mode, Full config}
    \label{fig:cpu-memory-full-pconn}
    \vspace{-3ex}
\end{figure*}

\subsubsection{Full Configuration}

In these experiments, we consider a number $N_i=100$ of configuration commands that needs to be sent by the SDN controller. 
In the persistent connection (`P-Conn') mode the $N_i$ commands are sent reusing one single TCP connection, in the non-persistent connection (`NP-Conn-Seq') mode a TCP connection is opened for each command, in the bulk scenario (`NP-Bulk'), a message which contains $N_i$ commands is prepared and sent to the SRv6 device (opening a new TCP connection). In all cases, we repeat the experiment $M_i=20$ times to evaluate the average response time. Figure~\ref{fig:full-insecure} shows the results of \textit{Full Config} experiments considering \textit{insecure} connection (only for gRPC and REST), while Figure~\ref{fig:full-secure} and~\ref{fig:full-secure-zoom} \textit{secure} connection (gRPC, REST, NETCONF and SSH/CLI). In all figures the error markers represent the 95\% Confidence Interval $CI_{95}$.

Considering the \textit{insecure} connection scenario, the performance (response time) of gRPC is better than REST for the `P-Conn' and `NP-Conn-Seq'.  In the `P-Conn' scenario we obtain for gRPC and REST a response time of around $0.20 s$ and $0.33 s$ respectively (for $N_i=100$ commands). It corresponds to 500 operations per second for gRPC and less than 300 op/s for REST, showing that in our implementation gRPC is more efficient than REST for the remotization of the configuration operations of the SRv6 device. Obviously, the achieved \textit{remote} control throughput is lower than the \textit{local} throughput of 1700 op/s achieved in the local rule enforcement (section \ref{sec:lre}). If the connection is established when sending each command (`NP-Conn-Seq') the control throughput further decreases to 300 op/s for gRPC and to 240 op/s for REST.

\begin{figure*}[!htbp]
    \centering
    \begin{subfigure}{0.62\columnwidth}
	    \centering
        \includegraphics[width=1\columnwidth]{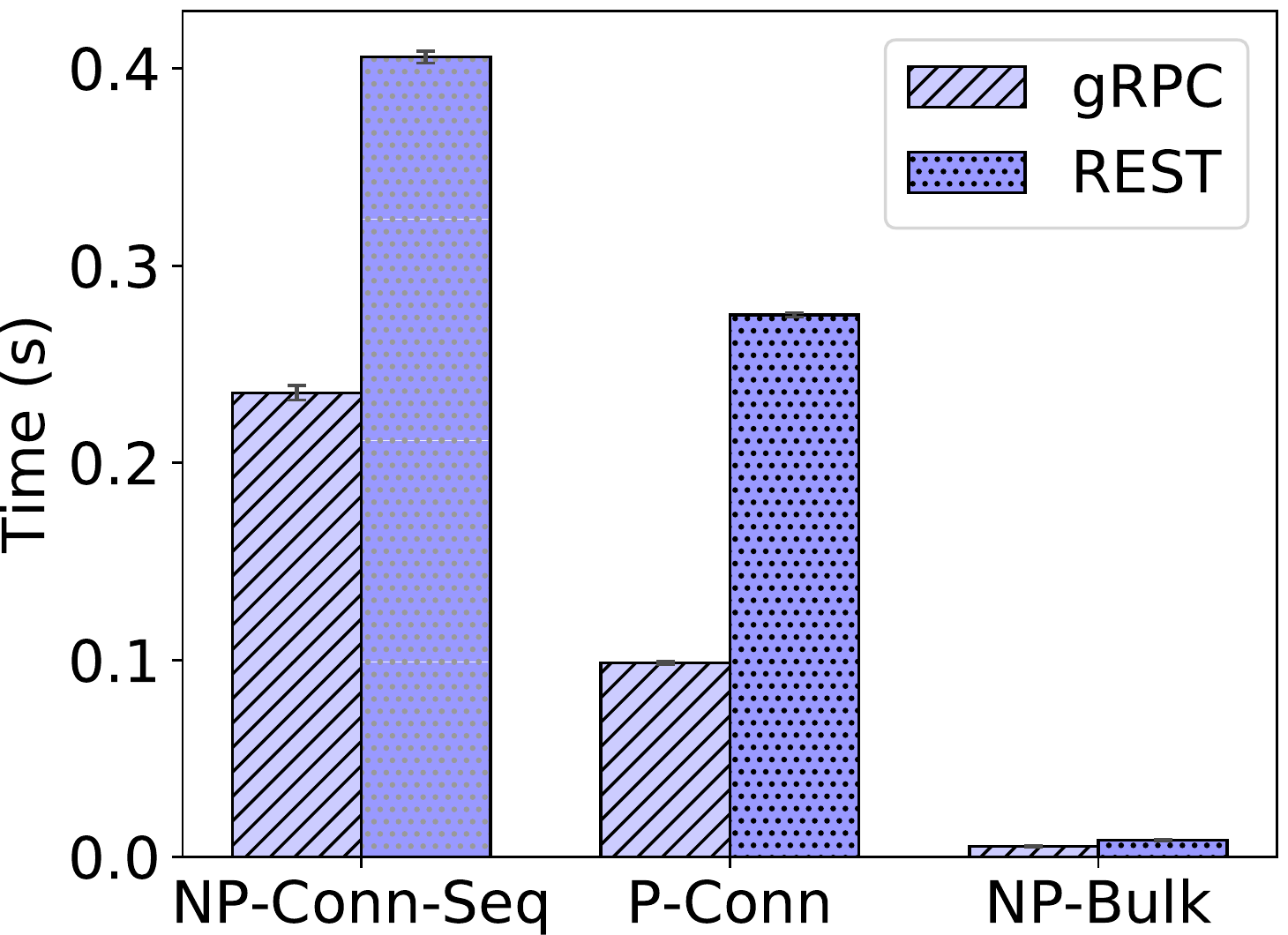}
        \caption{Insecure connections}
        \label{fig:communication-insecure}
    \end{subfigure}
    \begin{subfigure}{0.62\columnwidth}
	    \centering
        \includegraphics[width=1\columnwidth]{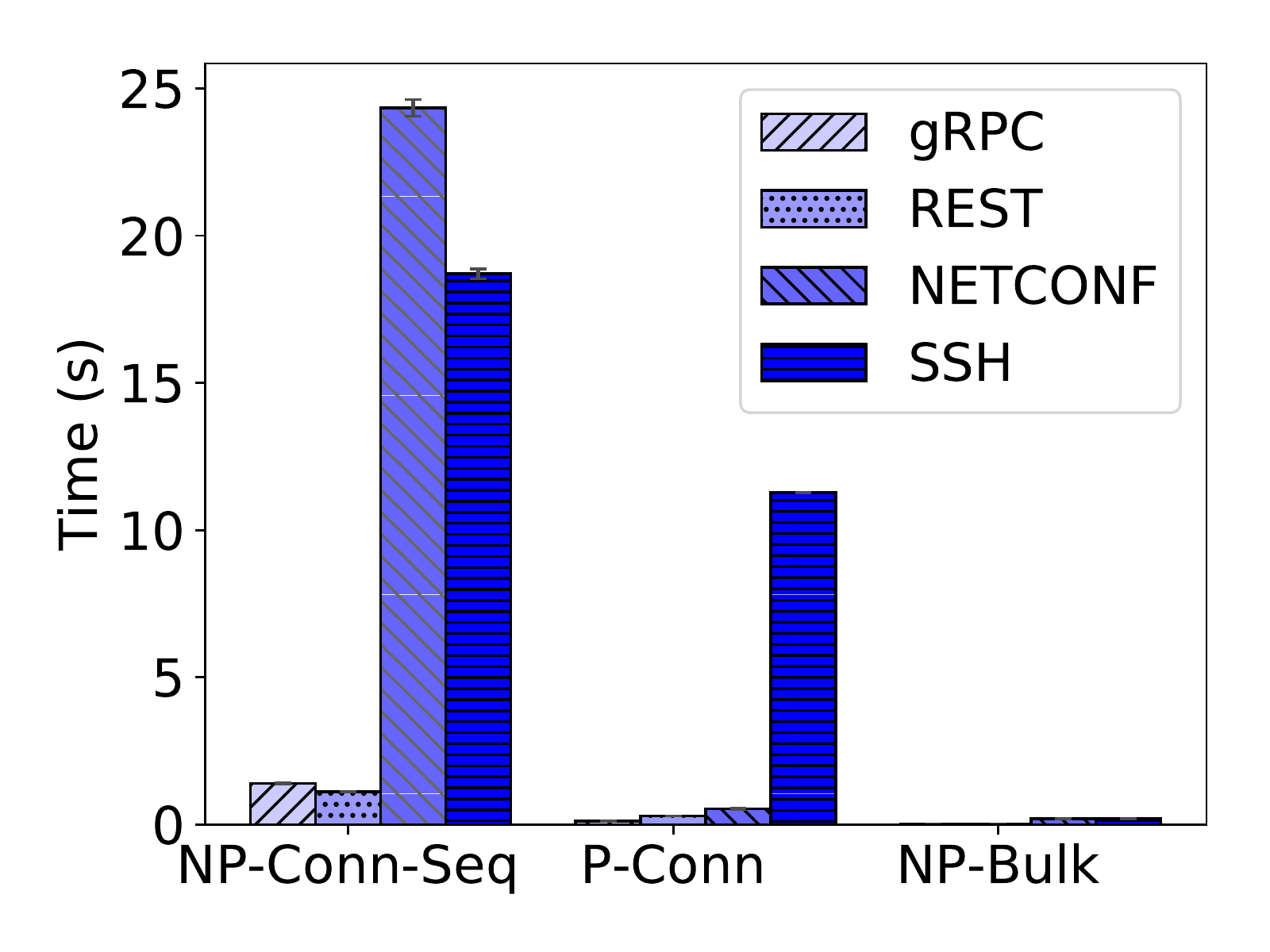}
        \caption{Secure connections}
        \label{fig:communication-secure}
    \end{subfigure}
    \begin{subfigure}{0.62\columnwidth}
	    \centering
        \includegraphics[width=1\columnwidth]{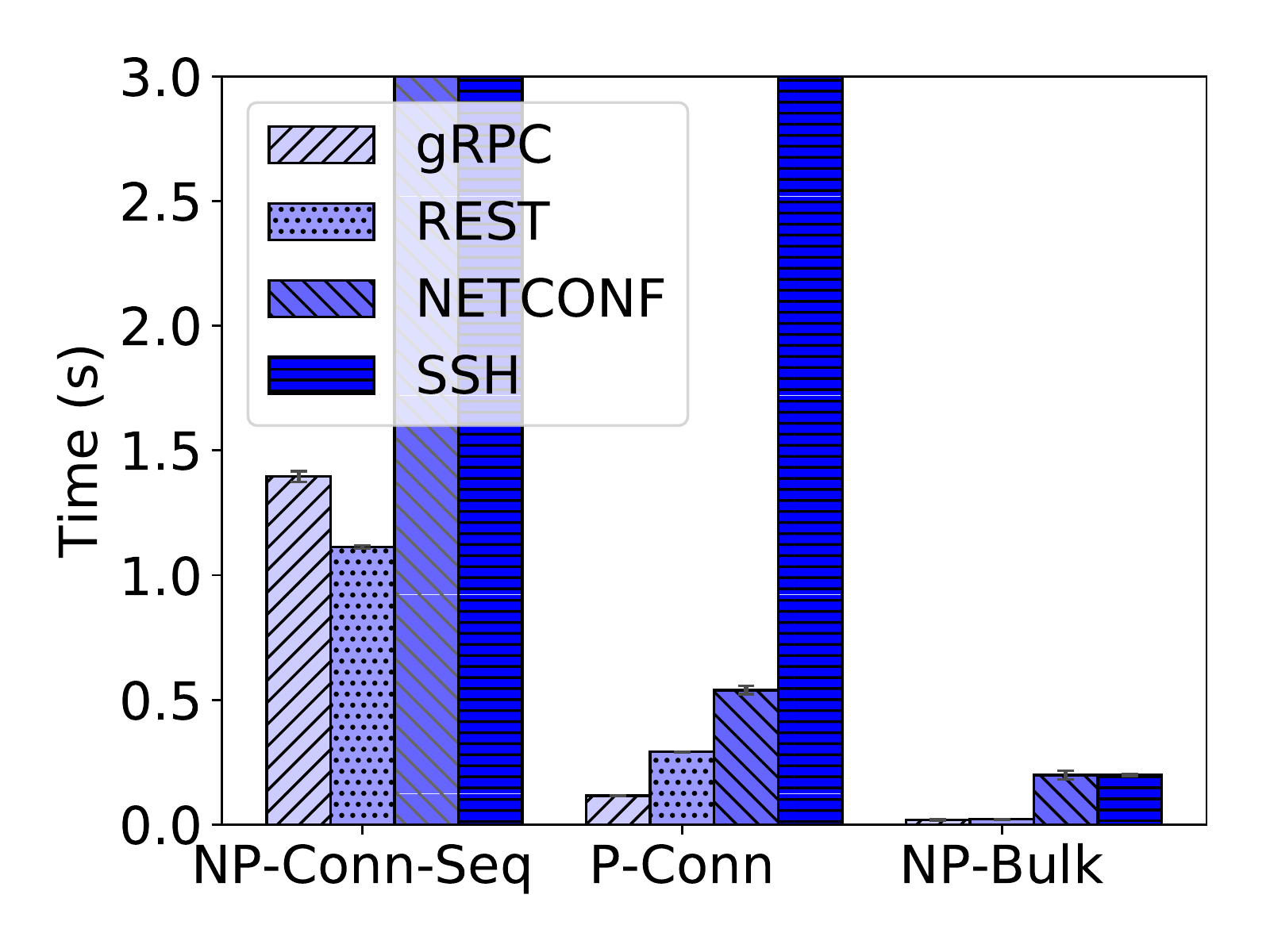}
        \caption{Secure connections (zoomed in)}
        \label{fig:communication-secure-zoom}
    \end{subfigure}
    \caption{Communication Only - Execution time}
    \label{fig:communication-response-time}
    \vspace{-3ex}
\end{figure*}

For the `NP-Bulk' insecure case, the response time for sending a single message with a 100 of commands is $0.087 s$ and $0.069 s$, respectively for gRPC and REST. As expected, these values are higher than the time needed to execute 100 local configurations ($0.065 s$), due to the time spent on the communication and message parsing parts. In this case REST is slightly better than gRPC. Fig.~\ref{fig:communication-response-time}(a)-NP-Bulk shows that for the communication part gRPC is faster than REST. Hence we conclude that the reason for the higher response time for gRPC is the parsing of the message content, which is implemented in a less efficient way in the gRPC case.

Using \textit{secure} connections it is possible to compare all the four implementations. From figure Figure~\ref{fig:full-secure} we see that in some cases in the `P-Conn' and `NP-Conn-Seq' scenarios the SSH/CLI and NETCONF have very poor performance (very high response time). For these reason we have added Figure~\ref{fig:full-secure-zoom} which focuses the y-axis in the range from 0 to 3 seconds, allowing to compare the performance of gRPC, REST and NETCONF for the `P-Conn' and `NP-Conn-Seq' scenarios. Looking at Figure~\ref{fig:full-secure-zoom} for the `P-Conn' case, the performance (response time) of NETCONF are worse than gRPC and REST, but in the same order of magnitude. In particular for gRPC, REST, NETCONF and SSH/CLI we obtain respectively a response time of $0.28$, $0.38$, $0.56$ and $12.2$ seconds (for 100 commands) which respectively correspond to a throughput of 357, 263, 178 and 8 operations per second. The very low performance of SSH/CLI is due to our poor implementation for this case, which reuses the TCP connection but creates a new secure socket for each command (i.e. the SSH authentication handshake is repeated each time). On the other hand NETCONF uses SSH as well, but performs the SSH handshake only at the beginning after setting up the TCP connection. To solve this issue, we should develop another version of the SSH/CLI implementation that behaves like NETCONF and this should drastically reduce the response time and increase the control throughput.

In the `NP-Conn-Seq' case, in which the connection is re-established at each command, the gRPC performance looks worse than REST. This is due to the initial TLS connection setup. Even if gRPC and REST use same protocol and same version, i.e. TLSv1.2, there are some differences in the initial handshake, gRPC uses more TLSv1.2 extensions. For this reason the setup phase is slower. Our empirical analysis showed that in general gRPC server replies after $0.073 s$, while REST after $0.049 s$. This slower setup phase has an impact in the `NP-Conn-Seq' case, while in the `P-Conn' case the slower setup phase is performed only once and the impact is negligible.

For gRPC and REST we can compare the results of the secure and insecure connections, as expected in the `P-Conn' scenario (persistent connection) there is a decrease in the throughput from 500 to 357 operations per second for gRPC and from 300 to 263 op/s for REST. With secure `NP-Conn-Seq' (non persistent), REST has better performance than gRPC. This means that the security setup phase of gRPC is slowing down its response time. The response time of NETCONF and SSH/CLI are very high ($23 s$ and $19 s$ respectively for 100 operations), because for each command the TCP connection is setup and then the SSH handshake is performed. Moreover, NETCONF introduces further overhead due to the creation of the NETCONF session. In `NP-Conn-Bulk' mode, only one message is sent (which contains 100 commands). gRPC and REST still show the smallest response time, $0.134 s$ and $0.126 s$ for 100 commands respectively. The NETCONF implementation has a response time of $0.321 s$ because of the SSH setup phase and of the setup of the NETCONF session.

As we have discussed for the local configuration case, we have performed CPU and memory measurements experiments in our comparison of the Southbound API implementations. The total system CPU load ($\%$) over time is reported Figures~\ref{fig:cpu-full-pc-grpcrestnetconf}~and~\ref{fig:cpu-full-pc-ssh}. The plots are obtained by requesting $100$ add commands using `P-Conn' mode (i.e. the commands are sent back-to-back over a single TCP connection). As already mentioned, considering that the system has 2 cores, the load of $50\%$ correspond to the full utilization of 1 core. Compared to the local configuration case, the gRPC and REST implementations show only a small decrease in the efficiency, as the CPU utilization is slightly less than the maximum load (due to the communication overhead, the CPU is less utilized). The communication overhead of NETCONF is higher, for this reason it takes more time to complete the execution of the commands and the average CPU utilization is lower. For the reasons that we have explained before, our SSH implementation is not efficient because of the re-establishment of a secure SSH socket for each command. This is clearly visible in Fig.~\ref{fig:cpu-full-pc-ssh}, which show a small CPU utilization for the long time interval needed to complete the execution of the commands (15 seconds). Finally, figures~\ref{fig:memory-full-pc-grpcrestnetconf} and~\ref{fig:memory-full-pc-ssh} show the memory usage on the device side for the different API implementations. The results confirm that the memory usage is relatively low in all cases, so it should not be a concern in practical cases.

\subsubsection{Communication Only}

In this part, we focus on the communication aspects of different implementations of the API. Therefore we implemented server side applications on the SRv6 device that accepts the requests from the client (SDN controller) and returns a predefined value without doing anything on the SRv6 device. Similar to the previous part, we make $N_i=100$ requests in the client side, repeat the experiment $M_i=20$ times and evaluate the average of overall response time. The results are shown in Fig.~\ref{fig:communication-response-time} and are consistent with the ones discussed in the previous subsection (\textit{Full config}), the differences among the API implementations are enhanced because no commands are actually executed in the SRv6 device. The gRPC implementation shows the best performance, followed by REST and NETCONF. The SSH/CLI implementation has a much worse performance, but we recall that we are repeating SSH authentication at each command, which reduces the performance of the `P-Conn' mode.

\begin{table}[!ht]
    \centering
    \begin{tabular}{|l|c|c||c|c|}\hline
    \multicolumn{5}{|c|}{\textbf{gRPC}}\\\hline
        &\multicolumn{2}{c||}{Insecure} & \multicolumn{2}{c|}{Secure}\\\cline{2-5}
        & Packets & Bytes & Packets & Bytes\\\hline
        Non-Persistent Con. &  1574 & 186 K & 1999 & 422 K \\\hline
        Persistent Con. & 216 & 35 K & 220 & 43 K \\\hline
        Bulk  & 21 & 6.5 K & 25 & 9 K\\\hline\hline
    \multicolumn{5}{|c|}{\textbf{REST}}\\\hline
        &\multicolumn{2}{c||}{Insecure} & \multicolumn{2}{c|}{Secure}\\\cline{2-5}
        & Packets & Bytes & Packets & Bytes\\\hline
        Non-Persistent Con.  &  1000 & 117 K & 1400 & 340 K \\\hline
        Persistent Con. & 208 & 63 K & 212 & 71 K \\\hline
        Bulk & 33 & 19 K  & 33 & 21 K\\\hline\hline
    \multicolumn{5}{|c|}{\textbf{SSH and NETCONF}}\\\hline
        &\multicolumn{2}{c||}{SSH} & \multicolumn{2}{c|}{NETCONF}\\\cline{2-5}
        & Packets & Bytes & Packets & Bytes\\\hline
        Non-Persistent Con. & 3107 & 493 K &  3648 & 743 K \\\hline
        Persistent Con. & 1020 & 130 K & 235 & 134 K \\\hline
        Bulk & 39 & 12 K & 70 & 36 K \\\hline
    \end{tabular}
    \caption{TCP Connection analysis}
    \label{tab:TCPConnections}
    \vspace{-2ex}
\end{table}

For further comparison of these APIs, we evaluated the transmitted packets and total transferred bytes\footnote{We captured the traffic over the communication link using tcpdump and analyzed it using Wireshark application.}. We send 100 add commands and report the observation in Table~\ref{tab:TCPConnections}. Also in terms of exchanged data, gRPC and REST are the most efficient solutions. gRPC has a larger overhead than REST in the initial setup phase, therefore gRPC sends more data than REST in the Non-Persistent connection case, while in Persistent connection and Bulk modes gRPC has the lowest overhead.

\subsubsection{Impact of Network Delay and Packet loss}

All the experiments reported so far have been run in ideal conditions, with the SDN controller close to the SRv6 device and connected through a LAN (or even a simple Ethernet cable). In this subsection we report a simple and obviously not exhaustive analysis of the impact of network delay and packet loss. In particular, we assume a fixed one way delay of 75 ms (which corresponds to 150 ms Round Trip Time (RTT) considering the two directions) and different packet loss ratios: 0\%, 0.5\%, 1\%, 2\% on each network interface. In the experiments we have synthetically applied the delay and loss ratio on the outgoing network interfaces of the SDN controller and of the SRv6 device using the \textit{netem} tool. The goal of this simple analysis is to verify if the introduction of these network impairments creates critical problems to our implementations of the Southbound API. 

\begin{figure}[!ht]
    \centering
    \begin{subfigure}{0.49\columnwidth}
        \centering
        \includegraphics[width=1\columnwidth]{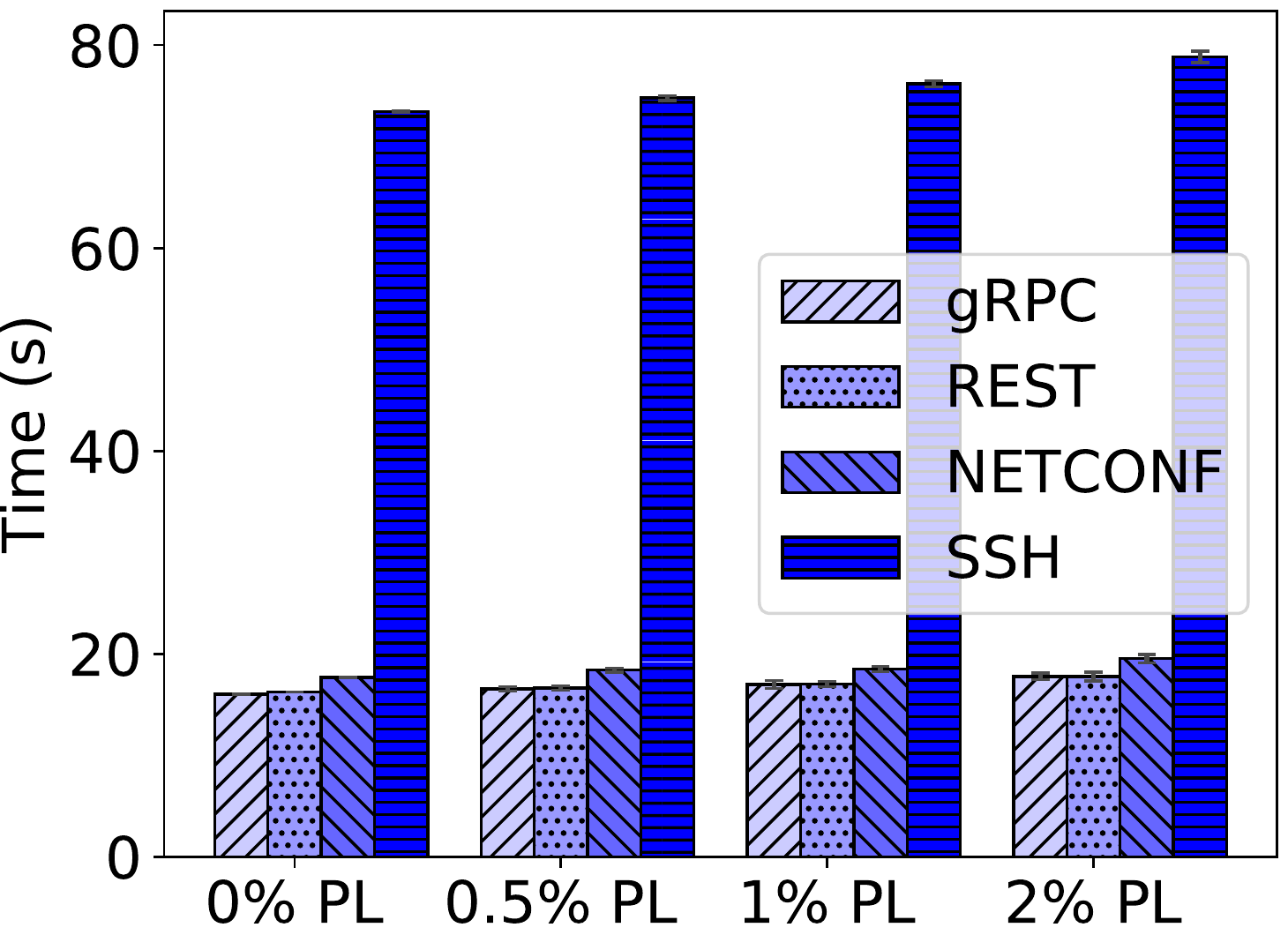}
        \caption{Secure connections}
    \end{subfigure}
    \begin{subfigure}{0.49\columnwidth}
        \centering
        \includegraphics[width=1\columnwidth]{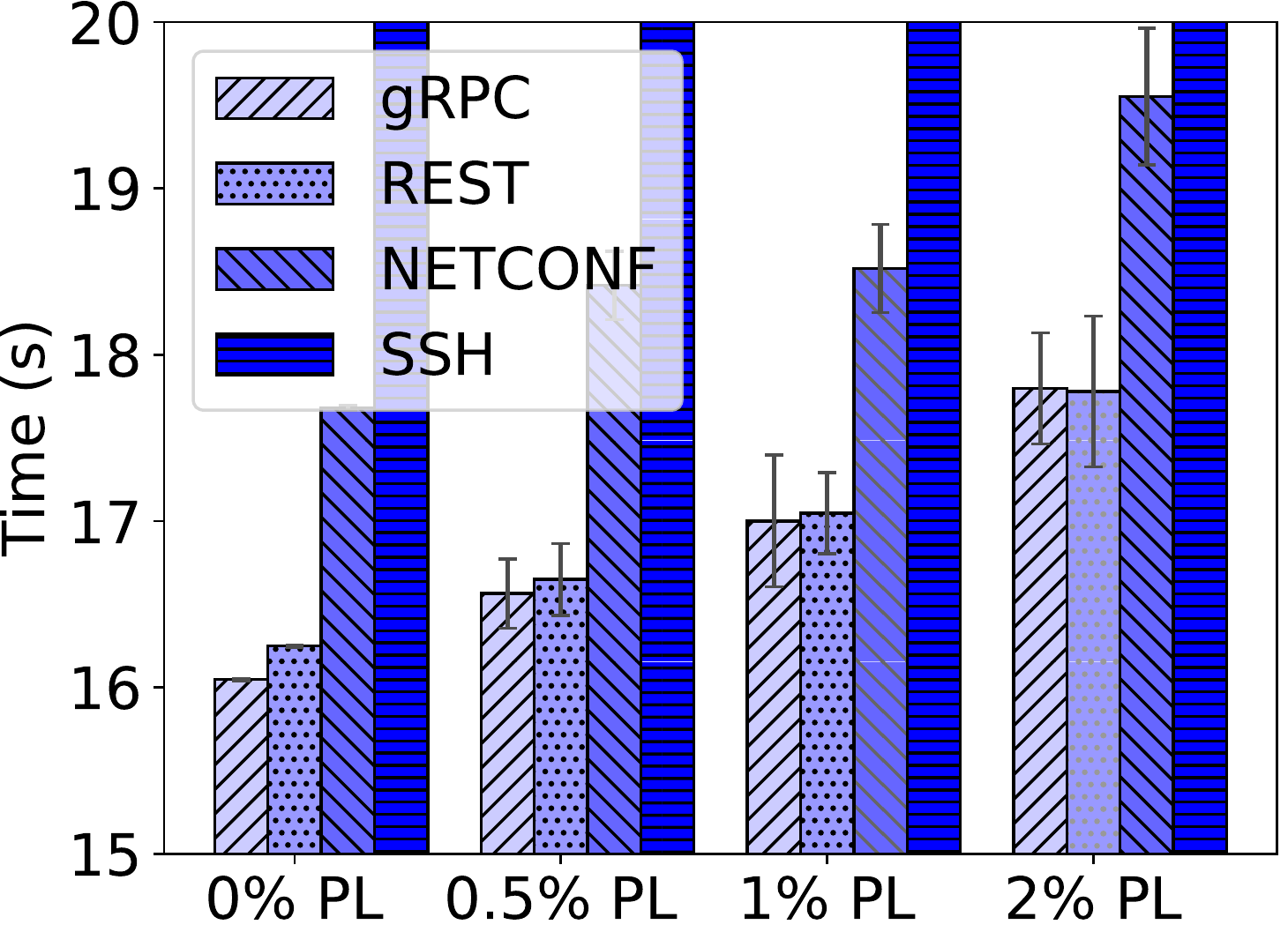}
        \caption{Secure connections (zoomed)}
    \end{subfigure}
    \caption{Response time of Full config in P-Conn scenario with secure connections and network impairments.} 
    \label{fig:full-response-time-insecure-nld}
    \vspace{-2ex}
\end{figure}

We assume that in real-world scenarios, the P-Conn mode (permanent connection) will be the most typical approach for the SDN controller to connect to switches and configure them. Therefore, in this section, we focus on the response time of P-Conn scenario. The response time of \textit{Full config} experiments are reported in Fig.~\ref{fig:full-response-time-insecure-nld}(a). Consider that every time a command is sent and the corresponding response is received by the controller, 150 ms of delay is added due to the Round Trip time. For 100 commands, the total time spent by packets traveling in the network is 15 seconds. Therefore in Fig.~\ref{fig:full-response-time-insecure-nld}(b) we plot only the part of the response time exceeding 15 seconds. When there is no loss the performance of gRPC and REST are good and similar each other (around 1 second is added to execute 100 commands), while NETCONF adds around 3 seconds to execute 100 commands. The SSH/CLI based mechanism is is not performing well in this case because the re-establishment of the secure socket for each command requires 3 additional Round Trip Times, so that 45 seconds are added to the minimum of 15 seconds needed to send the commands.

\subsubsection{Discussion of the results}

We found that gRPC and REST are the most efficient solutions, providing higher control throughput and lower response times. In the different scenarios that we have tested, these two implementations showed comparable performance. Despite gRPC being slightly better in most cases, we conclude that the choice between the two cannot be based on performance aspects. We plan to select gRPC for our future work as it offers a nice way to design, structure and manage the API thanks to the use of the Protocol Buffers which also drive the serialization of the data. Anyway this decision is rather subjective as the experiments show that the REST/HTTP 1.1 performance are comparable. The NETCONF implementation is less performant from the point of view of response time. Considering that the NETCONF approach provides more functionality, in particular offering inherent transaction capabilities, the performance loss that we have measured is still acceptable. For a network operator having devices already supporting NETCONF/Yang probably it will be better to re-use these technologies since the performance gains do not justify gRPC/REST approaches.

\subsection{Dynamic reconfiguration of SRv6 policies}
\label{sec:dynamic}

The possibility of dynamically change the network configuration in the devices is an important feature, this should happen with no impact on live traffic. For this reason, we have analyzed the effects of the dynamic configuration of SRv6 policies in the devices. In these experiments the effects of dynamic reconfiguration of SRv6 policies have been evaluated through the packet loss (to verify if a configuration change is hitless or not) and the traffic split. We define:

\begin{enumerate}
	\item Packet loss as the amount of packets lost during a network re-configuration;
	\item Traffic split refers to the distribution of the traffic due to network re-configurations.
\end{enumerate}

\begin{figure}[!t]
    \centering
    \includegraphics[trim={7cm 13cm 16cm 1.5cm},clip,width=0.32\textwidth]{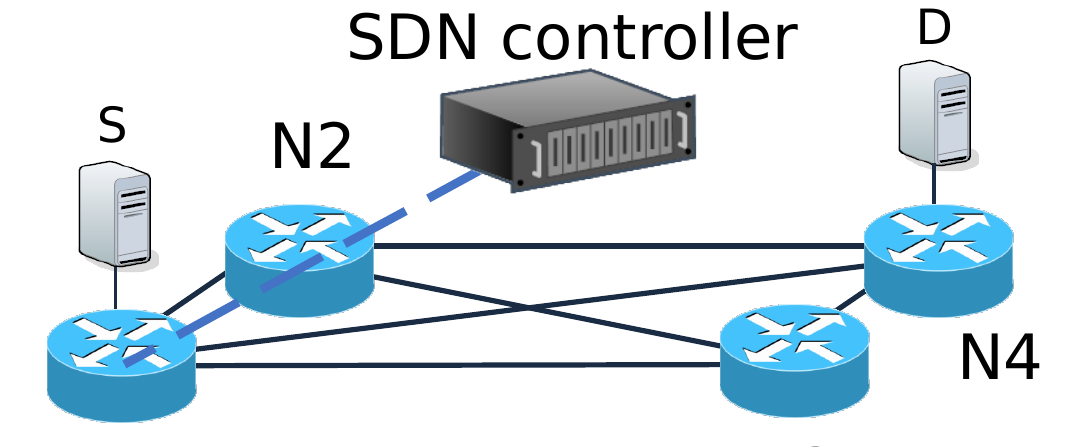}
    \caption{Dynamic reconfiguration experiment}
    \label{fig:exp_local_topo}
    \vspace{-3ex}
\end{figure}

We have analyzed the impact of dynamic reconfiguration of SRv6 policies in the SRv6 edge devices. In our physical testbed we have implemented the topology shown in Figure~\ref{fig:exp_local_topo}. The deployment is composed by four mini PCs with a low-energy Intel Celeron 1.3 Ghz dual core and 8GB of RAM. Each device is equipped with four Intel 82583V NICs at 1 Gb/s. One interface has been used as the management interface (not shown in Figure~\ref{fig:exp_local_topo}) and the other three for the direct interconnections with the other nodes (data plane network), so that a full mesh topology is realized. A recent Linux kernel (4.15) has been installed on the devices. In node N1 and N2 we have configured two network namespaces respectively acting as source (S) and destination (D) in our experiment. The SDN controller is running in an external node, connected via the management interface. In experiment described hereafter the controller is a Linux VM in a VirtualBox hypervisor running on a laptop.

We consider a flow of ICMPv6 packets (generated by the \textit{ping6} command) from the container \textit{S} in the node \textit{N1} to a container \textit{D} in the node \textit{N4}. Node \textit{N1} is acting as ingress edge device, while node \textit{N4} as egress device. The SDN controller is setting the SRv6 policies in the ingress device node \textit{N1}. In particular, the SDN controller associates the IPv6 destination address of the sink \textit{D} to three different SRv6 Segment lists: \textit{\{N4\}}, \textit{\{N2, N4\}}, \textit{\{N2, N3, N4\}}, each for a time interval of duration $T$ [s]. After that the SDN controller sets the first Segment List, the ICMPv6 flow is started by launching the \textit{ping6} application. Note that for each Segment list there is a different incoming interface in node \textit{N4}, so that we can count the SRv6 encapsulated packets incoming on the different interfaces to  easily evaluate how many packets have been transmitted using the three Segment lists. In the experiment the controller application uses the SSH/CLI approach to control the edge device. We also developed a local application running in the device (a bash script) that performs the same reconfiguration based on local timers. We refer to the experiments with the controller as \textit{remote configuration} and to the experiment with the application running in the edge device as \textit{local configuration}. 

\begin{table}[!ht]
\scalebox{1}{
    \begin{tabular}{|c|c|c|c|c|}
      \hline
         & 1s & 0.5s & 0.1s & 0.005s \\
      \hline
      REF   & $20,20,20$ & $40,40,40$ & $200,200,200$ & $4000,4000,4000$ \\
      \hline
      LOC   & $20,20,20$ & $40,40,40$ & $199,207,194$ & $4002,3997,4001$ \\
      \hline
      REM   & $20,20,20$ & $40,40,40$ & $198,207,195$ & $4003,4005,3992$ \\
      \hline
    \end{tabular}
    }
\centering 
\caption{Impact of the dynamic reconfiguration}
\label{tab:impact}
\vspace{-2ex}
\end{table}

We ran a number of experiments at different packet sending rates, by configuring the \textit{interval} parameter of the \textit{ping6} application, which defines the packet inter-departure interval [s]. First of all, we verified that the re-configurations are completely hitless, as no ICMPv6 lost packets have been reported by the \textit{ping6} application in any experiment, both in the local and in the remote experiments. By counting the packets on the incoming interfaces we verified that the distribution over the 3 Segments lists matches very well the configuration pattern. Table \ref{tab:impact} show the results of this test for a duration $T=20 [s]$.

In Table \ref{tab:impact} the row marked \textit{REF} represent the reference target values (i.e. evenly splitting the packets over the three Segment Lists), the \textit{LOC} and \textit{REM} rows refer respectively to the measured results of local configuration and remote configuration tests. The reported values represent the number of received packets over the three interfaces (i.e. on the three different segment lists). The columns represent different inter-departure packet intervals: 1s, 0.5s, 0.1s and 0.005s, which respectively correspond to a rate of 1, 2, 10 and 200 packets/s. As shown in Table \ref{tab:impact}, the results are pretty stable and show an optimal distribution of the packets both for local changes and remote configurations with slight differences from the reference target values.

\section{Conclusion}
\label{sec:conclusion}

In this paper, we have described an SDN based approach for controlling SRv6 enabled networks. We have discussed the architecture of SRv6 capable Linux nodes and the design of the SDN Southbound API offered to the controller. As regards this Southbound API, we have provided four implementations using different technologies (gRPC, REST, NETCONF, CLI/SSH). We released the implementation as Open Source and realized a testbed and a set of tools to easily replicate the proposed architecture and evaluate it with practical experiments. The different API implementations have been evaluated in terms of response time and CPU/memory utilization on the device. After the performance characterization, we concluded that in our implementation gRPC and REST show the best performance (in most cases gRPC is slightly better), NETCONF is less performant but still in the same order of magnitude (and it offers transaction capabilities). We conclude that there are no strong indications coming from the performance evaluation that could clearly drive the selection of one of these solutions for the Southbound API. The SSH/CLI implementation shows much lower performance, but we have identified a shortcoming in our design that could be addressed to improve the SSH/CLI performance. Finally, we have performed an analysis of the effects of the SRv6 configuration changes in the Linux SRv6 nodes, showing that we can achieve hitless reconfiguration of SRv6 policies with no packet loss.

\section*{Acknowledgment}
    This work has received funding from the Cisco University Research Program and from the EU H2020 5G-EVE project.
	
\bibliographystyle{IEEEtran}
\bibliography{main.bib}

\begin{thebibliography}{10}
\providecommand{\url}[1]{#1}
\csname url@samestyle\endcsname
\providecommand{\newblock}{\relax}
\providecommand{\bibinfo}[2]{#2}
\providecommand{\BIBentrySTDinterwordspacing}{\spaceskip=0pt\relax}
\providecommand{\BIBentryALTinterwordstretchfactor}{4}
\providecommand{\BIBentryALTinterwordspacing}{\spaceskip=\fontdimen2\font plus
\BIBentryALTinterwordstretchfactor\fontdimen3\font minus
  \fontdimen4\font\relax}
\providecommand{\BIBforeignlanguage}[2]{{%
\expandafter\ifx\csname l@#1\endcsname\relax
\typeout{** WARNING: IEEEtran.bst: No hyphenation pattern has been}%
\typeout{** loaded for the language `#1'. Using the pattern for}%
\typeout{** the default language instead.}%
\else
\language=\csname l@#1\endcsname
\fi
#2}}
\providecommand{\BIBdecl}{\relax}
\BIBdecl

\bibitem{idsrarch}
\BIBentryALTinterwordspacing
{\relax C. Filsfils and S. Previdi (eds.) et al.}, ``{Segment Routing
  Architecture},'' January 2018. [Online]. Available:
  \url{http://tools.ietf.org/html/draft-ietf-spring-segment-routing}
\BIBentrySTDinterwordspacing

\bibitem{filsfils2015segment}
C.~Filsfils \emph{et~al.}, ``{The Segment Routing Architecture},'' \emph{Global
  Communications Conference (GLOBECOM), 2015 IEEE}, pp. 1--6, 2015.

\bibitem{segment-routing.net}
\BIBentryALTinterwordspacing
{Segment Routing}. [Online]. Available: \url{http://www.segment-routing.net}
\BIBentrySTDinterwordspacing

\bibitem{idipv6srh}
\BIBentryALTinterwordspacing
{\relax S. Previdi (ed.) et al.}, ``{IPv6 Segment Routing Header (SRH)},''
  March 2018. [Online]. Available:
  \url{http://tools.ietf.org/html/draft-ietf-6man-segment-routing-header}
\BIBentrySTDinterwordspacing

\bibitem{SRv6blog-noction}
\BIBentryALTinterwordspacing
{Segment Routing and the SRv6 Network Programming}. [Online]. Available:
  \url{https://www.noction.com/blog/segment-routing-srv6-network-programming}
\BIBentrySTDinterwordspacing

\bibitem{fd-io-vpp}
``{What is VPP ?}'' \url{https://wiki.fd.io/view/VPP}.

\bibitem{mckeown2009software}
N.~McKeown, ``{Software-defined networking},'' \emph{INFOCOM keynote talk},
  vol.~17, no.~2, pp. 30--32, 2009.

\bibitem{kreutz2015software}
D.~Kreutz \emph{et~al.}, ``{Software-defined networking: A comprehensive
  survey},'' \emph{Proceedings of the IEEE}, vol. 103, no.~1, pp. 14--76, 2015.

\bibitem{rose}
\BIBentryALTinterwordspacing
{ROSE Project}. [Online]. Available: \url{https://netgroup.github.io/rose/}
\BIBentrySTDinterwordspacing

\bibitem{srv6-sdn}
\BIBentryALTinterwordspacing
{SRv6 SDN}. [Online]. Available: \url{https://netgroup.github.io/srv6-sdn/}
\BIBentrySTDinterwordspacing

\bibitem{mininet}
\BIBentryALTinterwordspacing
{Mininet Project}. [Online]. Available: \url{http://mininet.org}
\BIBentrySTDinterwordspacing

\bibitem{RFC7855}
{\relax S. Previdi et al.}, ``{Source Packet Routing in Networking (SPRING)
  Problem Statement and Requirements},'' IETF RFC 7855, May 2016.

\bibitem{RFC8355}
{\relax C. Filsfils et al.}, ``{Resiliency Use Cases in Source Packet Routing
  in Networking (SPRING) Networks},'' IETF RFC 8355, March 2018.

\bibitem{RFC8354}
{\relax J. Brzozowski et al.}, ``{Use Cases for IPv6 Source Packet Routing in
  Networking (SPRING)},'' IETF RFC 8354, March 2018.

\bibitem{id-ti-frr-sr}
\BIBentryALTinterwordspacing
A.~Bashandy \emph{et~al.}, ``{Topology Independent Fast Reroute using Segment
  Routing},'' March 2018. [Online]. Available:
  \url{http://tools.ietf.org/html/draft-bashandy-rtgwg-segment-routing-ti-lfa}
\BIBentrySTDinterwordspacing

\bibitem{interconnecting}
\BIBentryALTinterwordspacing
C.~Filsfils \emph{et~al.}, ``{Interconnecting Millions Of Endpoints With
  Segment Routing},'' Mar. 2018, work in Progress. [Online]. Available:
  \url{https://datatracker.ietf.org/doc/html/draft-filsfils-spring-large-scale-interconnect-09}
\BIBentrySTDinterwordspacing

\bibitem{srv6netprog}
\BIBentryALTinterwordspacing
C.~Fisfils \emph{et~al.}, ``{SRv6 Network Programming},'' March 2018. [Online].
  Available:
  \url{http://tools.ietf.org/html/draft-filsfils-spring-srv6-network-programming}
\BIBentrySTDinterwordspacing

\bibitem{salsano2016hybrid}
\relax S. Salsano~et al., ``{Hybrid IP/SDN networking: open implementation and
  experiment management tools},'' \emph{IEEE Transactions on Network and
  Service Management}, vol.~13, no.~1, pp. 138--153, 2016.

\bibitem{onl}
\BIBentryALTinterwordspacing
{Open Network Linux}. [Online]. Available: \url{https://opennetlinux.org}
\BIBentrySTDinterwordspacing

\bibitem{cumulus}
\BIBentryALTinterwordspacing
{Cumulus Networks}. [Online]. Available: \url{https://cumulusnetworks.com}
\BIBentrySTDinterwordspacing

\bibitem{rfc3549}
\BIBentryALTinterwordspacing
H.~M. Khosravi \emph{et~al.}, ``{Linux Netlink as an IP Services Protocol},''
  RFC 3549, Jul. 2003. [Online]. Available:
  \url{https://rfc-editor.org/rfc/rfc3549.txt}
\BIBentrySTDinterwordspacing

\bibitem{rfc5340}
\BIBentryALTinterwordspacing
D.~Ferguson \emph{et~al.}, ``{OSPF for IPv6},'' RFC 5340, Jul. 2008. [Online].
  Available: \url{https://rfc-editor.org/rfc/rfc5340.txt}
\BIBentrySTDinterwordspacing

\bibitem{rfc2545}
\BIBentryALTinterwordspacing
P.~R. Marques \emph{et~al.}, ``{Use of BGP-4 Multiprotocol Extensions for IPv6
  Inter-Domain Routing},'' RFC 2545, Mar. 1999. [Online]. Available:
  \url{https://rfc-editor.org/rfc/rfc2545.txt}
\BIBentrySTDinterwordspacing

\bibitem{lebrun2017implementing}
D.~Lebrun \emph{et~al.}, ``{Implementing IPv6 Segment Routing in the Linux
  Kernel},'' in \emph{Proceedings of the Applied Networking Research
  Workshop}.\hskip 1em plus 0.5em minus 0.4em\relax ACM, 2017, pp. 35--41.

\bibitem{pyroute2}
\BIBentryALTinterwordspacing
{Python Netlink library}. [Online]. Available:
  \url{https://pypi.python.org/pypi/pyroute2}
\BIBentrySTDinterwordspacing

\bibitem{quagga}
\BIBentryALTinterwordspacing
{Quagga project}. [Online]. Available: \url{https://www.quagga.net}
\BIBentrySTDinterwordspacing

\bibitem{ovs_sr}
``{OVS SRv6},'' \url{https://gitlab.flux.utah.edu/safeedge/ovs-srv6}, [Online;
  accessed 05-05-2017].

\bibitem{vissicchio2013safe}
S.~Vissicchio \emph{et~al.}, ``{Safe updates of hybrid SDN networks},''
  \emph{Universit{\'e} catholique de Louvain, Tech. Rep}, 2013.

\bibitem{mckeown2008openflow}
N.~McKeown \emph{et~al.}, ``{OpenFlow: enabling innovation in campus
  networks},'' \emph{ACM SIGCOMM Computer Communication Review}, vol.~38,
  no.~2, pp. 69--74, 2008.

\bibitem{rfc7047}
\BIBentryALTinterwordspacing
B.~Pfaff \emph{et~al.}, ``{The Open vSwitch Database Management Protocol},''
  RFC 7047, Dec. 2013. [Online]. Available:
  \url{https://rfc-editor.org/rfc/rfc7047.txt}
\BIBentrySTDinterwordspacing

\bibitem{smith-opflex-03}
\BIBentryALTinterwordspacing
M.~Smith \emph{et~al.}, ``{OpFlex Control Protocol},'' Internet Engineering
  Task Force, Internet-Draft draft-smith-opflex-03, Apr. 2016, work in
  Progress. [Online]. Available:
  \url{https://datatracker.ietf.org/doc/html/draft-smith-opflex-03}
\BIBentrySTDinterwordspacing

\bibitem{CiscoOnePK}
\BIBentryALTinterwordspacing
``{Cisco One Platform Kit (OnePK)},'' 2016. [Online]. Available:
  \url{https://www.cisco.com/c/en/us/products/collateral/ios-nx-os-software/bulletin-c25-736612.pdf}
\BIBentrySTDinterwordspacing

\bibitem{rfc5440}
\BIBentryALTinterwordspacing
J.~Vasseur \emph{et~al.}, ``{Path Computation Element (PCE) Communication
  Protocol (PCEP)},'' RFC 5440, Mar. 2009. [Online]. Available:
  \url{https://rfc-editor.org/rfc/rfc5440.txt}
\BIBentrySTDinterwordspacing

\bibitem{rfc6241}
\BIBentryALTinterwordspacing
R.~Enns \emph{et~al.}, ``{Network Configuration Protocol (NETCONF)},'' RFC
  6241, Jun. 2011. [Online]. Available:
  \url{https://rfc-editor.org/rfc/rfc6241.txt}
\BIBentrySTDinterwordspacing

\bibitem{rfc8040}
\BIBentryALTinterwordspacing
A.~Bierman \emph{et~al.}, ``{RESTCONF Protocol},'' RFC 8040, Jan. 2017.
  [Online]. Available: \url{https://rfc-editor.org/rfc/rfc8040.txt}
\BIBentrySTDinterwordspacing

\bibitem{cienawaveserver}
\BIBentryALTinterwordspacing
{Ciena}. {Ciena Waveserver}. [Online]. Available:
  \url{https://www.ciena.com/products/waveserver/}
\BIBentrySTDinterwordspacing

\bibitem{grpc}
\BIBentryALTinterwordspacing
{Google}. {GRPC: A high performance, open-source universal RPC framework}.
  [Online]. Available: \url{https://grpc.io/}
\BIBentrySTDinterwordspacing

\bibitem{protobuf}
\BIBentryALTinterwordspacing
{Google}. {Protocol Buffers}. [Online]. Available:
  \url{https://developers.google.com/protocol-buffers/}
\BIBentrySTDinterwordspacing

\bibitem{thrift}
\BIBentryALTinterwordspacing
{Apache Thrift}. [Online]. Available: \url{https://thrift.apache.org}
\BIBentrySTDinterwordspacing

\bibitem{fboss}
{Facebook}, ``{FBOSS Agent},'' \url{https://github.com/facebook/fboss},
  [Online; accessed 05-05-2017].

\bibitem{salsano2016pmsr}
S.~Salsano \emph{et~al.}, ``{PMSR - Poor Man's Segment Routing, a minimalistic
  approach to Segment Routing and a Traffic Engineering use case},'' in
  \emph{Network Operations and Management Symposium (NOMS), 2016
  IEEE/IFIP}.\hskip 1em plus 0.5em minus 0.4em\relax IEEE, 2016, pp. 598--604.

\bibitem{pang2017sdn}
{J. Pang and others}, ``{SDN-Based Data Center Networking With Collaboration of
  Multipath TCP and Segment Routing},'' \emph{IEEE Access}, vol.~5, pp.
  9764--9773, 2017.

\bibitem{sheu2017scalable}
J.~Sheu \emph{et~al.}, ``A scalable and bandwidth-efficient multicast algorithm
  based on segment routing in software-defined networking,'' in
  \emph{Communications (ICC), 2017 IEEE International Conference on}.\hskip 1em
  plus 0.5em minus 0.4em\relax IEEE, 2017, pp. 1--6.

\bibitem{davoli2015traffic}
L.~Davoli \emph{et~al.}, ``{Traffic engineering with segment routing: SDN-based
  architectural design and open source implementation},'' in \emph{Software
  Defined Networks (EWSDN), 2015 Fourth European Workshop on}.\hskip 1em plus
  0.5em minus 0.4em\relax IEEE, 2015, pp. 111--112.

\bibitem{srsurvey}
{P.L. Ventre and others}, ``{Segment Routing: A comprehensive survey of
  research activities, standardization efforts and implementation results},''
  \emph{arXiv preprint arXiv:?}, 2018.

\bibitem{stratum}
\BIBentryALTinterwordspacing
{ONF}. {Stratum project}. [Online]. Available: \url{https://stratumproject.org}
\BIBentrySTDinterwordspacing

\bibitem{OpenConfig}
\BIBentryALTinterwordspacing
``{OpenConfig Home Page},'' 2018. [Online]. Available:
  \url{http://www.openconfig.net/}
\BIBentrySTDinterwordspacing

\bibitem{tiesr}
\BIBentryALTinterwordspacing
{S. Salsano and others}. {Testbeds IntErconnections with L2 overlays - SRv6 for
  SFC}. [Online]. Available:
  \url{https://www.slideshare.net/stefanosalsano/testbeds-interconnections-with-l2-overlays-srv6-for-sfc}
\BIBentrySTDinterwordspacing

\bibitem{softfire}
\BIBentryALTinterwordspacing
{SoftFire Project}. [Online]. Available: \url{https://www.softfire.eu}
\BIBentrySTDinterwordspacing

\end{thebibliography}

\vspace{-4em}

\begin{IEEEbiography}[{\includegraphics[width=1in,height=1.25in,clip,keepaspectratio]{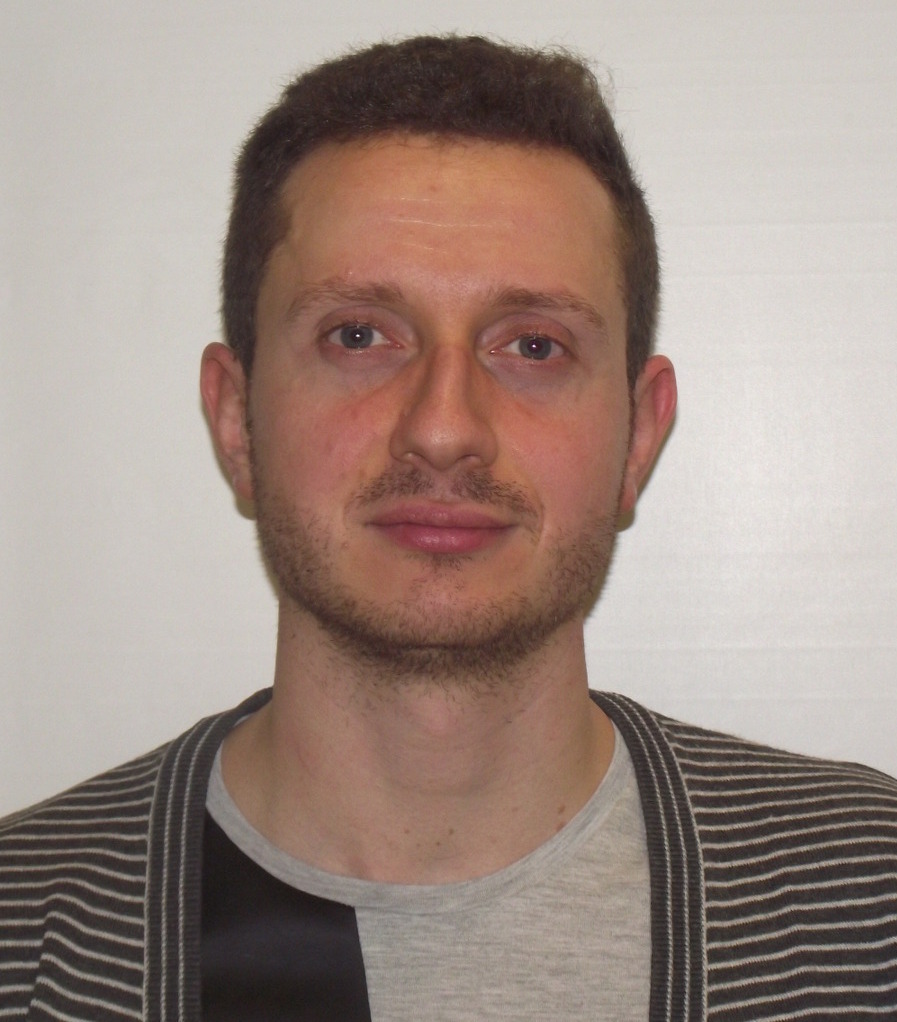}}]{Pier Luigi Ventre}
received his PhD in Electronics Engineering in 2018 from University of Rome ``Tor Vergata''. From 2013 to 2015, he was one of the beneficiary of the scholarship ``Orio Carlini'' granted by the Italian NREN GARR. His main research interests focus on Software Defined Networking, Network Function Virtualization, Virtualization and IPv6 Segment Routing. He worked as researcher in several projects founded by the EU and currently he is a post-doctoral researcher at CNIT.
\end{IEEEbiography}

\vspace{-4em}

\begin{IEEEbiography}[{\includegraphics[width=1in,height=1.25in,clip,keepaspectratio]{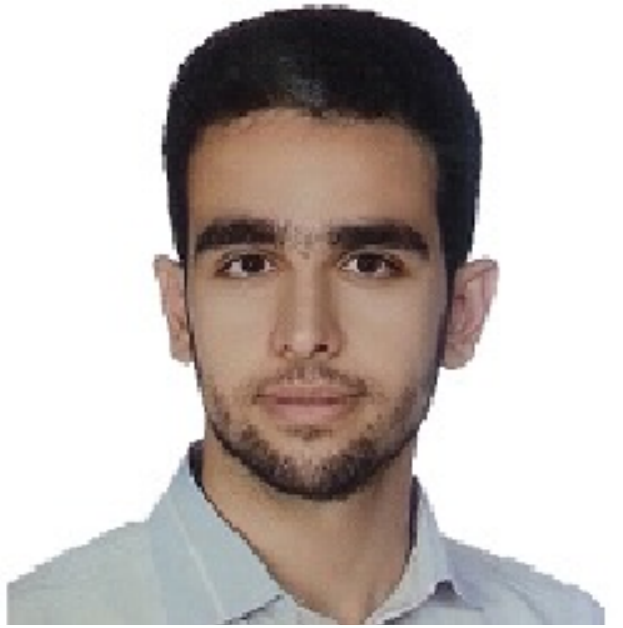}}]{Mohammad Mahdi Tajiki} graduated from Electrical and Computer Engineering School of Tehran University, Tehran, Iran. He is a PhD candidate in Tarbiat Modares University, Tehran, Iran. Currently, he is spending his sabbatical period in University Rome ``Tor Vergata''. His main research interests are Network QoS, media streaming over the Internet, data center networking, traffic engineering, and software defined networking (SDN). 
\end{IEEEbiography}

\vspace{-4em}

\begin{IEEEbiography}[{\includegraphics[width=1in,height=1.25in,clip,keepaspectratio]{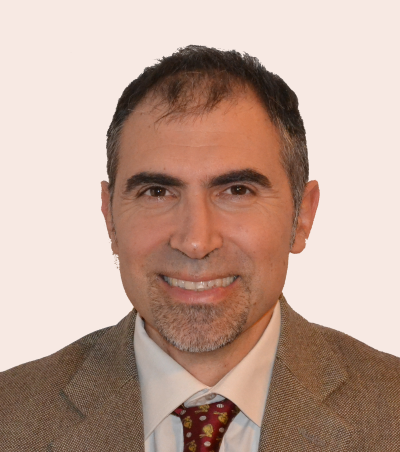}}]{Stefano Salsano}
(M'98-SM'13) received his PhD from University of Rome ``La Sapienza'' in 1998. He is Associate Professor at the University of Rome ``Tor Vergata''. He participated in 15 research projects founded by the EU, being project coordinator in one of them and technical coordinator in two. He has been PI in several research and technology transfer contracts funded by industries. 
His current research interests include SDN, Network Virtualization, Cybersecurity, Information Centric Networking.
\end{IEEEbiography}

\vspace{-4em}

\begin{IEEEbiography}[{\includegraphics[width=1in,height=1.25in,clip,keepaspectratio]{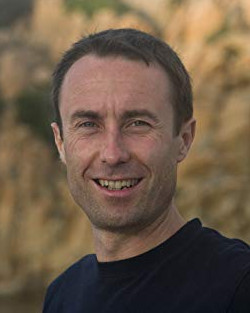}}]{Clarence Filsfils}
is a Cisco Systems Fellow, has a 20-year expertise leading innovation, productization, marketing and deployment for Cisco Systems. He invented the Segment Routing Technology and is leading its productization, marketing and deployment. Previously, he invented and led the Fast Routing Convergence Technology and was the lead designer for Cisco System's QoS deployments. Clarence is a regular speaker at leading industry conferences. He holds over 130 patents and is a prolific writer, either in academic circle, or standardization or books.
\end{IEEEbiography}

\end{document}